\documentclass[%
 pra,
 amsmath,amssymb,
 reprint,%
]{revtex4-1}

\usepackage{graphicx}
\usepackage{dcolumn}
\usepackage{bm}

\usepackage[utf8]{inputenc}
\usepackage[T1]{fontenc}
\usepackage{mathptmx}
\usepackage{etoolbox}
\usepackage[version=4]{mhchem}
\usepackage{stfloats}
\usepackage{graphicx}
\usepackage{subcaption}
\usepackage{float}
\usepackage{placeins}
\usepackage[justification=centering,singlelinecheck=false]{caption}

\makeatletter
\def\@email#1#2{%
 \endgroup
 \patchcmd{\titleblock@produce}
  {\frontmatter@RRAPformat}
  {\frontmatter@RRAPformat{\produce@RRAP{*#1\href{mailto:#2}{#2}}}\frontmatter@RRAPformat}
  {}{}
}%
\makeatother
\begin{document}

\preprint{AIP/123-QED}

\title{Time-dependent density functional theory study of strong-field laser-induced coulomb explosion of the HCl dimer}
\author{C. Jiang}
\affiliation{Department of Physics and Astronomy, Vanderbilt
University, Nashville,
Tennessee 37235, United States}
\author{Cody L. Covington}
\affiliation{ Department of Chemistry, Austin Peay State University,
Clarksville,
Tennessee 37044, United States}
\author{Kalman Varga}
\email{kalman.varga@vanderbilt.edu (corresponding author)}
\affiliation{Department of Physics and Astronomy, Vanderbilt
University, Nashville,
Tennessee 37235, United States}

\date{\today}

\begin{abstract}
We present a channel-resolved interpretation of laser-driven Coulomb explosion of 
the HCl dimer from an ensemble of trajectories. Three dominant outcomes are 
identified: a minor three-body channel and two four-body channels (sequential 
and near-simultaneous dissociation of both molecules). The key result is that 
pathway selection is strongly correlated with the degree of ionization during the laser 
interaction, which is in turn strongly modulated by laser-molecule orientation. 
Higher early-time ionization predisposes the system toward near-simultaneous 
four-body breakup, whereas lower ionization favors sequential and three-body 
fragmentation; for low-ionization cases, a fragment-resolved charge metric 
further differentiates three-body and sequential behavior. These charge-dependent 
trends consistently map onto experimentally accessible observables: the simultaneous 
mechanism dominates the high-energy tail of the kinetic energy release
(KER) spectrum and populate distinct regions of the 
emission-angle distributions, while sequential events concentrate 
at lower KER. Overall, early-time charge evolution provides a unifying 
explanation for channel branching and for the channel-resolved fragmentation signatures.
\end{abstract}

\maketitle

\section{\label{sec:level1}Introduction}

Coulomb explosion imaging (CEI) has emerged as a powerful route to “read out” molecular structure and dynamics by rapidly creating multiply charged states and recording the correlated momenta of the resulting fragment ions. When combined with laser-induced alignment and coincidence/covariance analyses, CEI can access stereochemical information in isolated molecules and weakly bound complexes, and—crucially—provide time-resolved structural observables on the femtosecond scale.\cite{Schouder2022CEIReview,Ashfold2023CEIPerspective}

In CEI, the measured fragment-ion momenta must be interpreted through a forward model that maps the explosion dynamics to the pre-fragmentation geometry.\cite{Vallance2019CEIForwardModel} Because strong-field/X-ray ionization, charge redistribution, and nuclear motion can overlap on femtosecond timescales, the common “instantaneous ionization + purely Coulombic repulsion” approximation can bias structural retrieval and channel identification if post-ionization dynamics are neglected.\cite{Slater2015PImMSCEI, Howard2023SlingshotEI} Therefore, modeling the dissociation dynamics is essential for quantitative CEI.

Coulomb explosion dynamics have been widely studied in recent years, spanning 
hydrocarbons\cite{Taylor2025PRA1, Li2020EthaneNSDI, Mogyorosi2025ButaneCE}, 
carbonyl compound\cite{Zhao2021OCSCEI,Ren2003COCEI, Tseng2018H2COCEI}, 
halogenated hydrocarbon\cite{Itsukashi2018DiiodoCE, Wu2019BrClCE, Trost2025CH2I2}, 
molecular cluster\cite{Zhang2024NH3ClusterCE,Zhao2025HClDimer, Mery2021CODimerCEI, Schouder2020CS2Dimer}.
Considerable effort has been directed toward characterizing the
dissociation dynamics and fragmentation pathways of small molecular
systems exposed to high-intensity laser pulses or energetic electron
impact
\cite{
doi:10.1021/acs.jpca.3c05442,10.1063/5.0117875,
PhysRevA.104.053104,PhysRevA.110.053104,s7v8-5hmk,jjbn-vtjm,r6c2-jzh3,6h8l-dyxc,
Voigtsberger2014,PhysRevLett.120.113202,PhysRevA.107.043115,PhysRevLett.129.023001,
PhysRevA.105.063105,PhysRevA.104.053104,doi:10.1021/acs.jpclett.2c01007,
PhysRevLett.130.233001,7f9s1d4t,Venkatachalam2025,Stamm2025,
Severt2024}.

Zhao \textit{et al.} \cite{Zhao2025HClDimer} 
used COLTRIMS to study four-body Coulomb explosion of the HCl dimer in the coincidence channel $\mathrm{H^+ + H^+ + Cl^+ + Cl^+}$ under a strong-field femtosecond laser
($800~\mathrm{nm}$, $35~\mathrm{fs}$).
Their analysis centers on the kinetic energy release and ion--ion angular/energy correlations, showing that 
the H-H emission angle is broadly distributed while the Cl-Cl emisssion
angle is more confined, and that the proton-emission geometry can strongly influence the energy partitioning of the two $\mathrm{Cl^+}$ fragments. These trends are interpreted as arising from the intrinsic asymmetry of the neutral dimer, together with a separation of timescales between rapid proton separation and a slower, Coulomb-driven Cl--Cl breakup, supported by
ab initio molecular dynamics 
AIMD simulations.\cite{Zhao2025HClDimer} 

These measurements provide an essential macroscopic picture of the
breakup kinematics; however, a complementary, channel-resolved
microscopic interpretation---linking distinct dissociation pathways to
their underlying charge evolution and nuclear motion is not studied.
Motivated by this gap, we address the following question: how do different fragmentation channels emerge from the coupled electron--nuclear dynamics during a laser-driven Coulomb explosion of the HCl dimer, and what early-time signatures predispose the pathway selection?
To address this, we perform a channel-resolved mechanistic analysis of HCl-dimer dissociation by linking the time-dependent structural evolution in simulated trajectories to experimentally measurable long-time observables, such as the KER and emission-angle distributions. Instead of assuming a fixed initial charge state, we explicitly follow ionization during the laser pulse and quantify the early-time total and fragment charges, together with the relative orientation between the laser polarization and the dimer/bond axes. This allows us to identify systematic correlations between early-time ionization and subsequent channel branching, and to interpret how pathway selectivity is reflected in the energy- and angle-resolved fragmentation signatures.

To capture this interplay directly, we employ real-time time-dependent
density functional theory
(TDDFT)\cite{PhysRevLett.52.997,Ullrich2011TDDFTbook} on a real-space grid as our main theoretical tool. In contrast to approaches that rely on precomputed potential-energy surfaces, real-time TDDFT propagates electrons and nuclei together under a time-dependent Kohn--Sham Hamiltonian, thereby capturing the coupled interaction among electrons, nuclei, and the laser field, as well as charge migration, state mixing, and energy flow on attosecond–femtosecond timescales.\cite{ComNano,Attosecond} This approach provides direct access to the time-resolved electron-density distribution of the system, which is central to the present work, enabling us to relate early-time ionization-and its orientation dependence during the laser interaction-to subsequent channel branching and, ultimately, to the long-time KER and emission-angle signatures. This method has already delivered accurate ultrafast dynamics across diverse systems.\cite{Russakoff2015PRA,Russakoff2015PRA2,Covington2017PRA,Jiang2025JCP,Taylor2025PRA1,Taylor2025PRA2,Bubin2012PRA,Wang2025}

\section{\label{sec:level2}Computational Method}

The molecular dynamics in each set of simulations are modeled by real-time
time-dependent density-functional theory (TDDFT) on a real-space grid.  
The Kohn–Sham Hamiltonian has the following form:

\begin{equation}
\begin{split}
\hat H_{\mathrm{KS}}(t)
= {}& -\frac{\hbar^{2}}{2m}\nabla^{2}
     +V_{\mathrm{ion}}(\mathbf r,t)
     +V_{H}[\rho](\mathbf r,t) \\
  & +V_{XC}[\rho](\mathbf r,t)
     +V_{\mathrm{laser}}(\mathbf r,t).
\end{split}
\tag{1}
\end{equation}
where the first term, $-\hbar^{2}\nabla^{2}/2m$, is the single-electron kinetic-energy
operator.  
The electron density is  
\begin{equation}
\rho(\mathbf r,t)=
   \sum_{k=1}^{N_{\text{orbitals}}}
   2\bigl|\psi_{k}(\mathbf r,t)\bigr|^{2},
\tag{2}
\end{equation}
calculated by summing the contributions from all occupied Kohn–Sham orbitals.
$V_{\mathrm{ion}}$ denotes the external potential due to the ions, modeled with
norm-conserving pseudopotentials centered on each nucleus as given by Troullier
and Martins \cite{Troullier1991PRB}.
$V_{H}$ is the Hartree potential, representing the electrostatic Coulomb
interaction between the electrons,
\begin{equation}
V_{H}(\mathbf r,t)=
  \int \frac{\rho(\mathbf r^{\prime},t)}
            {|\mathbf r-\mathbf r^{\prime}|}\,
        d\mathbf r^{\prime},
\tag{3}
\end{equation}
The exchange-correlation potential V$_{XC}$ is 
approximated using the generalized gradient approximation (GGA), 
developed by Perdew et al. \cite{Perdew1981PRB}.
The last term in Eq.~(1), $V_{\mathrm{laser}}(\mathbf r,t)$, is the time-dependent potential induced by the laser electric field.
Within the dipole approximation, it is written as $V_{\mathrm{laser}}(\mathbf r,t)=\mathbf r\cdot\mathbf E_{\mathrm{laser}}(t)$.The electric field $\mathbf E_{\mathrm{laser}}(t)$ is taken to be 
\begin{equation}
\mathbf E_{\mathrm{laser}}(t)
= E_{\max}\exp\!\left[-\frac{(t-t_{0})^{2}}{2a^{2}}\right]\sin(\omega t)\,\hat{\mathbf k},
\tag{4}
\end{equation} 
where $E_{\max}$, $t_{0}$, and $a$ specify the peak amplitude, the temporal center of the pulse, and the width of the Gaussian envelope, respectively.
The parameter $\omega$ denotes the laser frequency, and $\hat{\mathbf k}$ is the unit vector along the polarization direction of the electric field.

Before the time-dependent calculations, we perform a
density-functional-theory (DFT) calculation to obtain the ground state of the
system, including the equilibrium electron density, self-consistent Kohn–Sham
orbitals, and the total energy.  With these initial conditions, we propagate the
orbitals in time using the time-dependent Kohn–Sham equation
\begin{equation}
i\hbar
\frac{\partial\psi_{k}(\mathbf r,t)}{\partial t}
      =\hat H_{\mathrm{KS}}(t)\,\psi_{k}(\mathbf r,t),
\tag{5}
\end{equation}
which is integrated with the propagator
\begin{equation}
\psi_{k}(\mathbf r,t+\delta t)=
   \exp\!\Bigl[-\,\tfrac{i}{\hbar}\hat H_{\mathrm{KS}}(t)\,\delta t\Bigr]
   \psi_{k}(\mathbf r,t)
\tag{6}
\end{equation}
and approximated by a fourth-order Taylor expansion:
\begin{equation}
\psi_{k}(\mathbf r,t+\delta t)\approx
   \sum_{n=0}^{4}\frac{1}{n!}
   \bigl(-\tfrac{i\delta t}{\hbar}\hat H_{\mathrm{KS}}(t)\bigr)^{n}
   \psi_{k}(\mathbf r,t).
\tag{7}
\end{equation}
The orbitals are propagated for $N$ time steps up to
$t_{\mathrm{final}} = N\Delta t$, where the time step $\Delta t$ is 
1~attosecond. This small timestep guarantees a conditionally stable time-propagation preserving the norm of the orbitals.
The $t_{\mathrm{final}}$ is set to 120 femtosecond, which is sufficient for the dissociation and fragmentation dynamics of interest to occur in the vast majority of cases.

A uniform $100\times100\times100$ cubic grid with a spacing of
0.30~Å (an $30.0\times30.0\times30.0$~Å$^{3}$ box) represents the orbitals in
real space.  A 1~as time step and a 0.30~Å grid spacing have been shown to yield
accurate results in previous dissociation dynamics studies.\cite{Taylor2025PRA1,Taylor2025PRA2,Jiang2025JCP}

The Kohn–Sham orbitals are set to zero at the box boundaries.  To prevent
unphysical reflections of electron density when fragments reach the edge, a complex absorbing
potential (CAP) surrounds the box.  Our simulations employ the form proposed by
Manolopoulos\cite{Manolopoulos2002JCP}:
\begin{equation}
-iw(x)= -\,\frac{i\hbar^{2}}{2m}
        \Bigl(\frac{2\pi}{\Delta x}\Bigr)^{2}
        f(y),\qquad
y=\frac{x-x_{1}}{\Delta x},
\tag{8}
\end{equation}
where $x_{1}$ and $x_{2}$ are the start and end of the absorbing region,
$\Delta x=x_{2}-x_{1}$, $c=2.62$ is a numerical constant, $m$ is the electron
mass, and
\begin{equation}
f(y)=\frac{4}{c^{2}}
      \left(\frac{1}{(1+y)^{2}}
            +\frac{1}{(1-y)^{2}}-2\right).
\tag{9}
\end{equation}

Ionic motion is treated classically within the Ehrenfest approximation.  The
electron-induced forces are
\begin{equation}
M_{i}\frac{d^{2}\mathbf R_{i}}{dt^{2}}=
  \sum_{j\neq i}^{N_{\text{ion}}}
      \frac{Z_{i}Z_{j}(\mathbf R_{i}-\mathbf R_{j})}
           {|\mathbf R_{i}-\mathbf R_{j}|^{3}}
  \;-\;
  \nabla_{\mathbf R_{i}}
  \int V_{\mathrm{ion}}(\mathbf r;\mathbf R_{i})\,
       \rho(\mathbf r,t)\,d\mathbf r,
\tag{10}
\end{equation}
where $M_{i}$, $Z_{i}$, and $\mathbf R_{i}$ are the mass, pseudocharge
(valence), and position of ion $i$, respectively, and
$N_{\text{ion}}$ is the total number of ions.  Equation~(10) is integrated with
the Verlet algorithm at every time step $\Delta t$.

\begin{figure}[H]
  \centering
  \includegraphics[width=0.4\textwidth]{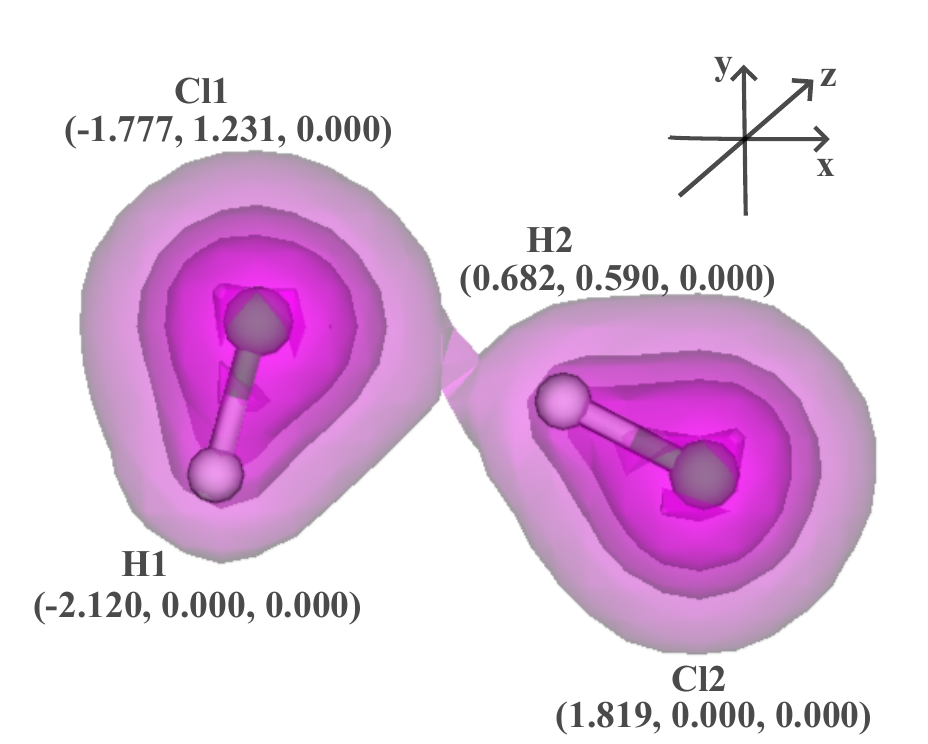}
  \caption{Ground-state geometry and electron-density distribution of the HCl dimer. The Cartesian coordinates are given in angstrom (\AA). Four linearly spaced electron-density isosurfaces are shown. The isosurface values shown are 0.10, 0.567, 1.033, and 1.50.}
  \label{fig1}
\end{figure}

The initial (HCl)$_2$ structure was taken directly from the reported equilibrium (global-minimum) geometry in Ref.~\cite{Xue2025HXDimers} 
The dimer was then placed at the center of the simulation box to maximize the available simulation space.
Fig.~\ref{fig1} shows the ground-state geometry and electron-density distribution of the HCl dimer.
For consistency, we label the four atoms as H1, H2, Cl1, and Cl2 throughout the paper.
Accordingly, the two monomers are denoted as HCl1 (H1--Cl1) and HCl2 (H2--Cl2).

\begin{figure}[H]
  \centering
 \includegraphics[width=0.5\textwidth]{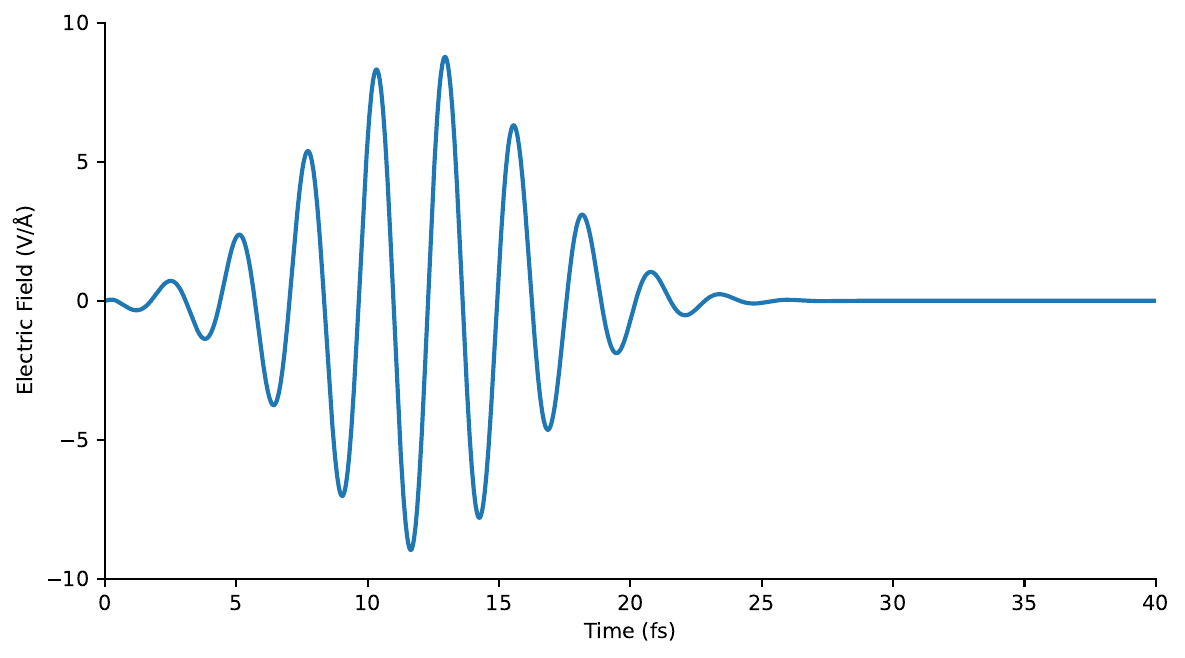}
  \caption{Electric field (unit: V/\AA) versus time (unit: fs) for the laser applied in the simulations.}
  \label{fig2}
\end{figure}

We apply a linearly polarized near-infrared few-cycle laser pulse with a central wavelength of
790~nm and a pulse duration of 6~fs (FWHM). The peak field strength is set to
$E_{\max}=9~\mathrm{V/}$\text{\AA}. Fig. \ref{fig2} shows the electric field of this laser from 0 fs to 40 fs. 
The applied laser electric field is turned on at $t=0$~fs, reaches its maximum amplitude at
$t \approx 12\text{--}13$~fs, and then decays to near zero by $t \approx 25$~fs. These parameters are chosen to target an ensemble-averaged net
ionization of $\sim 4$ electrons (i.e., an effective charge state $q \approx 4$), corresponding to
the charge window where four-body breakup is observed experimentally.\cite{Zhao2025HClDimer} 
These laser parameters closely match the experimental values for pulse
duration, frequency, and intensity; however, the exact experimental
pulse shape is not known, and our analytic form should be regarded as
an approximation.\cite{Zhao2025HClDimer}
We therefore examine how the ionization level relates to the channel branching within this experimentally relevant charge regime.

To mimic an isotropic gas-phase ensemble, the polarization direction $\hat{\mathbf k}$ is randomized independently for each trajectory.
Specifically, $\hat{\mathbf k}$ is sampled uniformly on the unit sphere by drawing $\phi\in[0,2\pi)$ uniformly and $u=\cos\theta\in[-1,1]$ uniformly, and setting $\hat{\mathbf k}=(\sqrt{1-u^{2}}\cos\phi,\sqrt{1-u^{2}}\sin\phi,u)$.
Moreover, initial ionic velocities are sampled randomly from a Boltzmann distribution at
300~K to simulate the random nuclear motions.  This approach allows diverse fragmentation channels and provides a
holistic view of the fragmentation dynamics. A total of 96 simulations were performed in this study.

Our computational results yield fractional electron ejection within the system, particularly during ionization processes. This behavior admits multiple interpretations. One perspective suggests that a specific fraction of electrons maintains localization within the molecular domain throughout the simulation timeframe, where the fractional charge may either undergo recombination with the ionized electron distribution or undergo dissociation over extended simulation periods. A more pragmatic interpretation views the non-integer charge as a statistical average: individual molecular fragments may preserve varying numbers of valence electrons, with some retaining one electron count while others maintain different values.

In the following section, data including time for H-Cl bond breaking, kinetic energy release (KER), H-H emission angle, Cl-Cl emission angle, and HCl fragment charge will be given. We present below a definition for each of these quantities.

\paragraph{H--Cl bond-breaking time.}
For each H--Cl pair, the bond-breaking time $t_\mathrm{br}$ is defined as the first time at which the
internuclear distance reaches $R_{\mathrm{HCl}}(t_{\mathrm{br}})\ge 2.0~\text{\AA}$ and remains
$R_{\mathrm{HCl}}(t)\ge 2.0~\text{\AA}$ for the subsequent 20~fs.
This threshold is chosen to be far beyond the equilibrium H--Cl bond of 1.274 \AA~and the typical vibrational excursion,
thereby suppressing false positives from transient bond stretching/recrossing and providing a robust, operational
criterion for irreversible dissociation on the simulation timescale. We define $t_{\mathrm{br},1}$ and $t_{\mathrm{br},2}$ to be the H1--Cl1 and H2--Cl2 bond-breaking times, respectively.

\paragraph{Kinetic energy release (KER).}
The total kinetic energy release is evaluated at the final time as
\begin{equation}
\mathrm{KER}=\sum_{i\in\{\mathrm{H1,H2,Cl1,Cl2}\}} \frac{1}{2} m_i \left|\mathbf{v}_i\right|^2,
\tag{11}
\end{equation}
where $\mathbf{v}_i$ is obtained by a backward finite difference using the last two nuclear frames,
$\mathbf{v}_i \approx \big(\mathbf{r}_i(t_{\mathrm{last}})-\mathbf{r}_i(t_{\mathrm{last}}-\Delta t)\big)/\Delta t$ with $\Delta t=0.5$ fs,
and $m_i$ are the nuclear masses (H and Cl). The individual fragment kinetic energies and their sum are reported in eV.

\paragraph{H--H and Cl--Cl emission angles.}
The H--H emission angle $\theta_{\mathrm{HH}}$ (and analogously $\theta_{\mathrm{ClCl}}$) is defined from the angle between the
asymptotic emission directions of the two corresponding nuclei. Operationally, we estimate each emission direction by
averaging the velocity vectors over the last 3 steps (forming two backward finite-difference velocities) to reduce numerical noise:
\begin{equation}
\bar{\mathbf{v}}_i
=
\frac{1}{2}
\sum_{k=k_f-1}^{k_f}
\mathbf{v}_i^{(k)} ,
\qquad
\theta_{\mathrm{HH}}=\cos^{-1}\!\left(\frac{\bar{\mathbf{v}}_{\mathrm{H1}}\cdot \bar{\mathbf{v}}_{\mathrm{H2}}}
{|\bar{\mathbf{v}}_{\mathrm{H1}}|\,|\bar{\mathbf{v}}_{\mathrm{H2}}|}\right),
\tag{12}
\end{equation}
with an identical definition for $\theta_{\mathrm{ClCl}}$ using Cl1 and Cl2. Because momentum is proportional to velocity
in the non-relativistic regime, these angles correspond to momentum-space emission angles used in fragmentation analyses.

\paragraph{HCl fragment charge.}
To quantify the charge localized on each HCl molecule, we define a fragment electron number by integrating the electronic density
within a sphere of radius $R=2.0$~\text{\AA} centered at the H--Cl bond midpoint $\mathbf{r}_0$:
\begin{equation}
N_e^{\mathrm{(HCl)}}(R)=\int_{|\mathbf{r}-\mathbf{r}_0|\le R}\rho(\mathbf{r})\,d^3\mathbf{r}.
\tag{13}
\end{equation}
The corresponding fragment charge is then
\begin{equation}
q_{\mathrm{HCl}}(R)=8-N_e^{\mathrm{(HCl)}}(R),
\tag{14}
\end{equation}
where 8 is the valence electron count of neutral HCl. We have tested $R=1.8$, 2.0, and 2.2~\text{\AA}; while the absolute values of
$q_{\mathrm{HCl}}$ shift with $R$, the qualitative channel-resolved conclusions remain unchanged. The radius-dependence test is
summarized in the Supporting Information.

In the next section, we present the channel-resolved dissociation dynamics of the (HCl)$_2$ dimer under the applied few-cycle laser pulse. Specifically, we (i) summarize the channel branching ratios together with the corresponding ionization statistics, (ii) analyze the total and fragment-resolved charges at 13 fs to identify early-time signatures correlated with each channel, (iii) quantify how the alignment of the laser polarization with the dimer center-of-mass(COM) axis and with the two H–Cl bond axes correlates with channel selectivity, (iv) show representative trajectory snapshots for all major dissociation pathways, including their time-resolved nuclear motion correlated with the laser field and total ionization, and (v) compare the KER and H–H/Cl–Cl angular distributions of the four-body channels against the reference experiment. 

\section{\label{sec:level3}Results and Discussion}

\subsection{\label{sec:level4}Channel classification and ionization–branching correlation}

Three major fragmentation pathways are observed across the 96 trajectories. 
Throughout this work, all times are reported as absolute simulation times measured from the start of the TDDFT propagation ($t=0$). 
In all trajectories, the H1--Cl1 bond breaks first, with
$t_{\mathrm{br,1}}\in[14.5,\,22.5]$~fs, after which the dynamics branch into three channels:

(a) \textit{Three-body dissociation:} the H2--Cl2 bond remains intact, showing only oscillation, leading to the breakup
\ce{H} + \ce{Cl} + \ce{HCl}.

(b) \textit{Four-body simultaneous dissociation:} the H2--Cl2 bond also breaks within a comparable time window to H1--Cl1, yielding the four-body products
\ce{H} + \ce{Cl} + \ce{H} + \ce{Cl}.

(c) \textit{Four-body sequential dissociation:} HCl2 persists as a metastable intermediate and the H2--Cl2 bond breaks at a later \emph{absolute} time,
$t_{\mathrm{br,2}}\in[22.0,\,79.5]$~fs (relative to $t=0$, not relative to the first bond dissociation), ultimately producing
\ce{H} + \ce{Cl} + \ce{H} + \ce{Cl}.

For four-body events we define the inter-break time
\begin{equation}
\Delta t \equiv t_{\mathrm{br},2}-t_{\mathrm{br},1},
\tag{15}
\end{equation}
where $t_{\mathrm{br},1}$ and $t_{\mathrm{br},2}$ are the earlier and later H--Cl bond-breaking times, respectively.
We classify a trajectory as \emph{simultaneous} if $\Delta t \le 6~\mathrm{fs}$ and as \emph{sequential} otherwise.

This choice is motivated by (i) the pulse duration ($\sim 6~\mathrm{fs}$ FWHM), such that two bond dissociations within this window
are naturally associated with the same strong-field ionization burst, (ii) the empirical $\Delta t$ histogram, which shows
a low-count transition region on the few-to-ten femtosecond scale separating the dominant near-zero peak from delayed events
(see Supporting Information), and (iii) nuclear timescales: on sub-$\sim$10~fs delays (comparable to or shorter than one H--Cl
vibrational period), nuclear rearrangement is expected to remain limited, whereas for longer delays appreciable bond stretching/rotation
and Coulomb-driven nuclear motion can develop, consistent with a metastable intermediate preceding the second bond breaking.

\begin{figure}[H]
  \centering
  \includegraphics[width=0.5\textwidth]{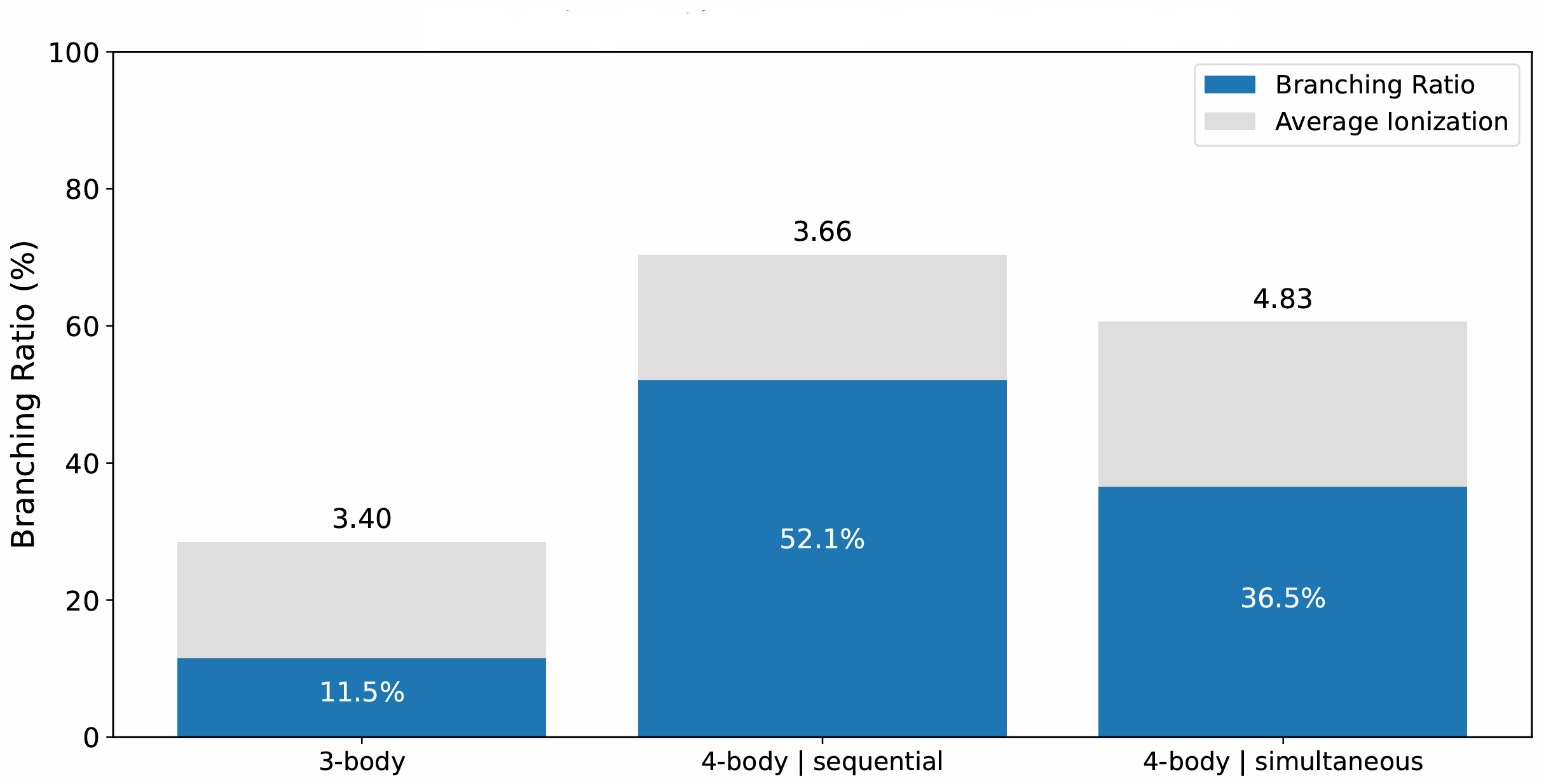}
  \caption{Branching ratio and average total ionization(at 25fs, when the laser field has essentially vanished) for each of the three dissociation channels obtained from 96 TDDFT trajectories. The average total ionization across all trajectories is 4.06\,e.}
  \label{fig3}
\end{figure}

We obtained the branching ratios in Fig.~\ref{fig3} based on the above channel classification. Four-body breakup dominates the ensemble (85/96 trajectories), whereas three-body breakup is relatively rare (11/96). Within the four-body events, the sequential pathway (50/96) is more prevalent than the simultaneous pathway (35/96). The charge-resolved analysis 
reveals that the simultaneous four-body channel is associated with a substantially larger average total ionization (4.83\,e) than either the three-body (3.40\,e) or the sequential four-body (3.66\,e) channel.

Notably, the mean total ionization differs across the three channels,
suggesting that the pathway selection may be strongly predisposed by
the early-time ionization. This observation motivates a
charge-resolved analysis as follows.

We choose t=13 fs as a representative early-time snapshot within the ionization window but prior to the first H–Cl bond breaking (which is H1-Cl1 in all trajectories), so that the extracted charge distributions reflect laser-induced ionization rather than later charge redistribution during large-amplitude nuclear separation.

\begin{figure}[H]
  \centering
\includegraphics[width=0.5\textwidth]{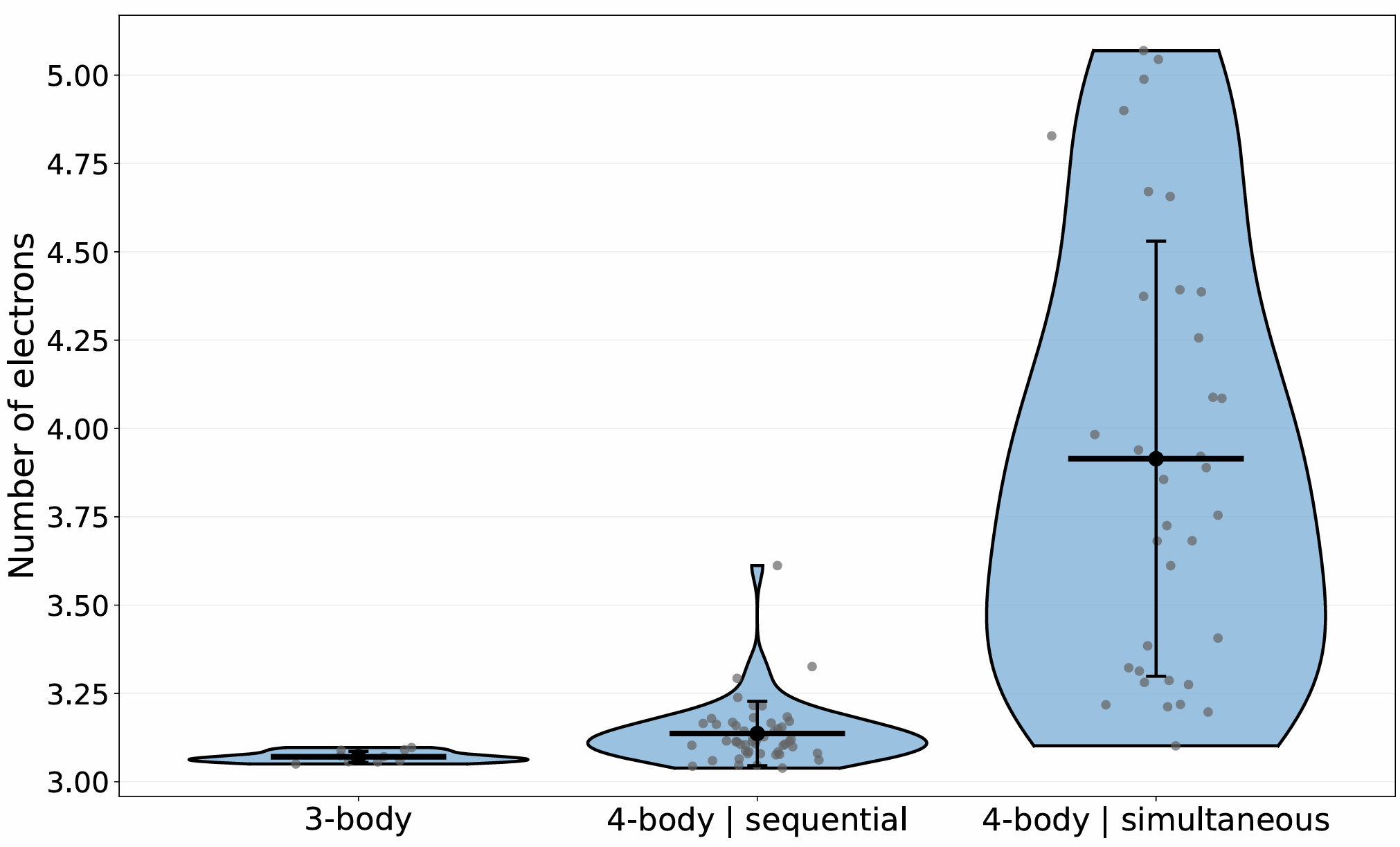}
  \caption{Violin plot of the total ionization at $t=13$~fs grouped by dissociation channel. The mean total ionization (mean $\pm$ standard deviation) is $3.07 \pm 0.02~e$ for 3-body events, $3.14 \pm 0.09~e$ for 4-body sequential events, and $3.91 \pm 0.62~e$ for 4-body simultaneous events.}
  \label{fig4}
\end{figure}

\begin{figure}[H]
  \centering
\includegraphics[width=0.5\textwidth]{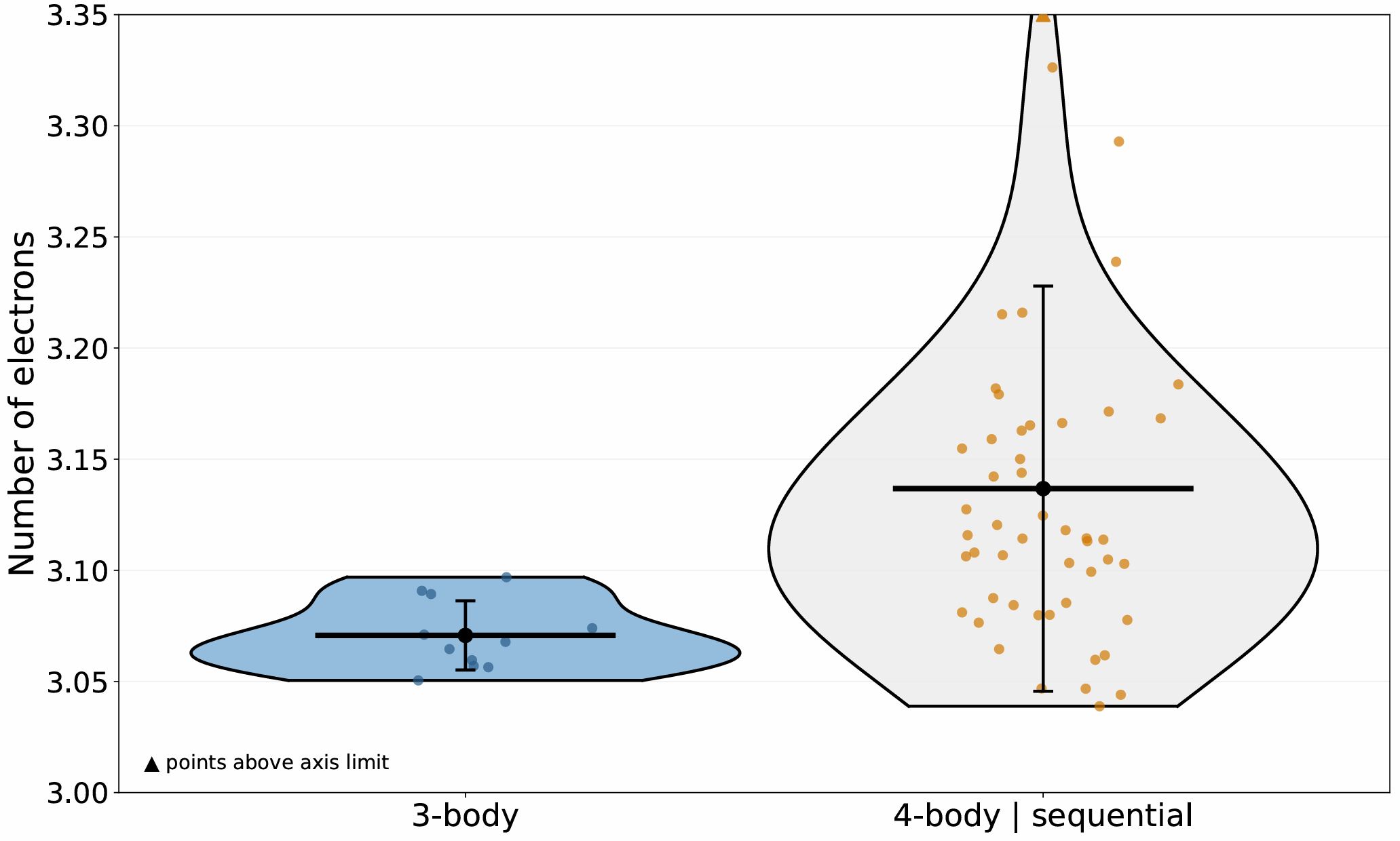}
  \caption{Zoomed-in view of the total ionization distributions for 3-body and 4-body sequential events, shown with a reduced $y$-axis range to resolve the low-ionization region.}
  \label{fig5}
\end{figure}

Fig.~\ref{fig4} and Fig.~\ref{fig5} summarize the total ionization distributions at 13~fs for the three dissociation channels.
For 3-body events, the total ionization is narrowly clustered around $\sim$3.1 electrons with a very small spread.
For 4-body sequential events, the distribution remains relatively tight and centers near $\sim$3.15 electrons.
In contrast, 4-body simultaneous events exhibit a much broader distribution, spanning from $\sim$3.2 to $>5$ electrons, with a mean of 3.91 electrons. Only a small fraction of the lowest-ionization simultaneous events overlaps with the highest-ionization tail of the sequential distribution.

Overall, the global net-ionization metric provides substantial separation between the two four-body pathways, suggesting that the overall degree of ionization is strongly correlated with whether four-body breakup proceeds via the simultaneous or sequential mechanism. By comparison, the 3-body and 4-body sequential channels show similar ensemble-averaged total ionization. Since their primary mechanistic difference is whether HCl2 dissociates, we next examine the fragment charge of HCl2 to assess whether a local, fragment-resolved charge metric can further distinguish these two channels.

\begin{figure}[H]
  \centering
\includegraphics[width=0.5\textwidth]{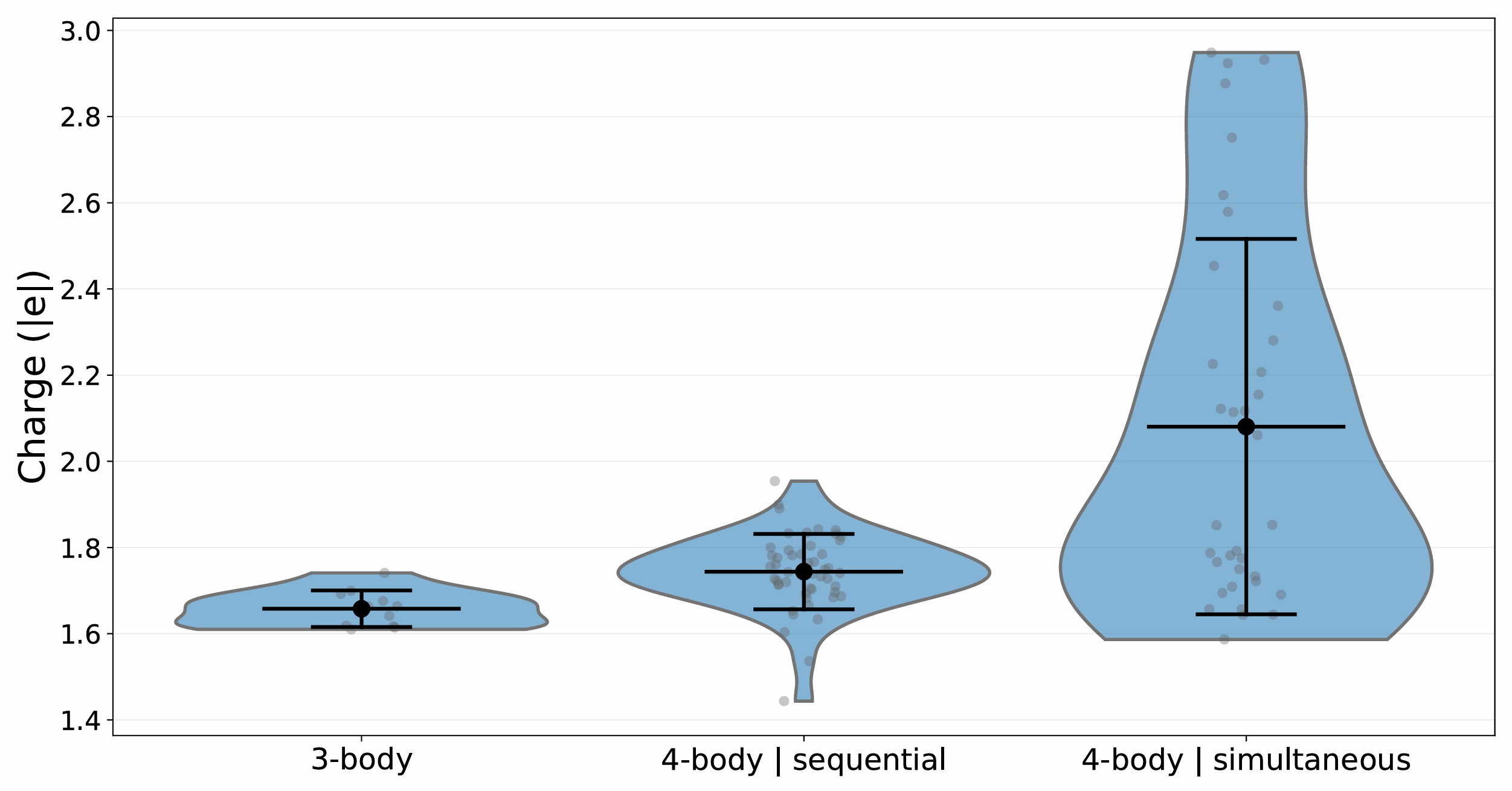}
  \caption{Violin plot of the HCl2 fragment charge at $t=13$~fs grouped by dissociation channel. The mean charge (mean $\pm$ standard deviation) is $1.66 \pm 0.04~e$ for 3-body events, $1.74 \pm 0.09~e$ for 4-body sequential events, and $2.08 \pm 0.44~e$ for 4-body simultaneous events.}
  \label{fig6}
\end{figure}

\begin{figure}[H]
  \centering
\includegraphics[width=0.5\textwidth]{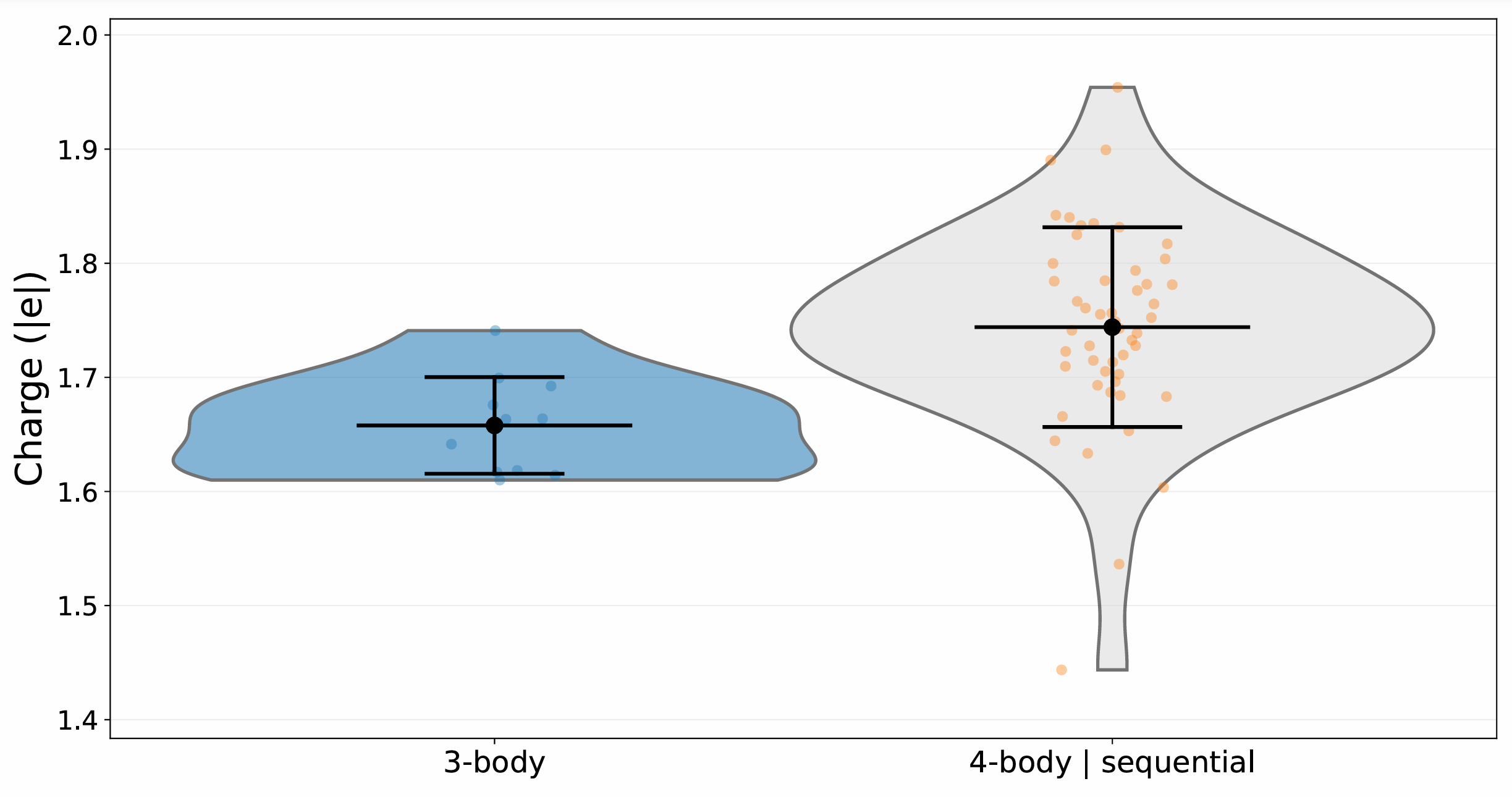}
  \caption{Zoomed-in view of the HCl2 fragment charge distributions for 3-body and 4-body sequential events, shown with a reduced $y$-axis range to resolve the low-ionization region.}
  \label{fig7}
\end{figure}

Fig.~\ref{fig6} and Fig.~\ref{fig7} summarize the HCl2 fragment-charge distributions at $t=13$~fs for 
the three dissociation channels using violin plots. 
The violin plot displays individual data points with slight random
horizontal displacement to prevent overlapping, while maintaining
their actual vertical positions.
As shown in Fig.~\ref{fig6}, 3-body events exhibit a narrowly peaked HCl2 charge centered at $1.66~e$ with a small spread, whereas 4-body sequential events are systematically shifted to a slightly higher mean value ($1.74~e$) with a broader distribution.
The 4-body simultaneous channel shows the largest mean charge on HCl2 ($2.08~e$) and the widest variability, spanning from $\sim 1.6~e$ to above $2.9~e$, consistent with substantially stronger and more heterogeneous ionization in this pathway.
To resolve the subtle but systematic separation between the low-ionization channels, Fig.~\ref{fig7} provides a zoomed-in view comparing 3-body and 4-body sequential events; despite partial overlap, the sequential distribution is clearly skewed toward higher HCl2 charge and has a higher-density peak than the 3-body distribution.

Notably, the mean HCl2 charge difference between 3-body and 4-body sequential events ($\Delta q_{\mathrm{HCl2}}\approx 0.08~e$) is comparable to their mean difference in total ionization at 13~fs ($\Delta q_{\mathrm{tot}}\approx 0.07~e$).
This correspondence suggests that the extra ionization distinguishing the sequential channel is primarily localized on the HCl2 molecule, rather than being distributed uniformly across the dimer.
Although the absolute shift is modest, it is systematic and therefore provides a useful indicator for separating 3-body and 4-body sequential pathways.

Taken together, Figs.~\ref{fig3}--\ref{fig7} establish a clear correlation between early-time ionization and the ensuing dissociation pathway.
At the global level, the total ionization at $t=13$~fs already provides strong separation between the two four-body mechanisms: simultaneous events exhibit substantially higher and much more heterogeneous ionization than sequential events, with only minor overlap at the distribution tails.
At the local level, while the total ionization is similar for the 3-body and sequential channels, the HCl2 fragment charge shows a small but systematic upward shift for sequential events.
This indicates that the additional ionization distinguishing the sequential channel is preferentially localized on the HCl2 molecule, which is consistent with its fragmentation.

\subsection{\label{sec:level5}Laser–molecule orientation analysis}

Given that early-time ionization strongly correlates with the ensuing dissociation pathway, 
we next investigate the origin of this ionization variability. Because the laser polarization 
is randomized over an fixed initial dimer geometry, the trajectory ensemble naturally samples 
different laser–molecule orientations. We therefore quantify how the ionization at 25~fs 
(when the laser field has essentially vanished) depends on the polarization–axis alignment with the 
dimer COM axis and with each HCl bond axis, and evaluate whether this orientation dependence provides a mechanistic link between the laser interaction and channel selectivity.

\begin{figure}[H]
  \centering
\includegraphics[width=0.5\textwidth]{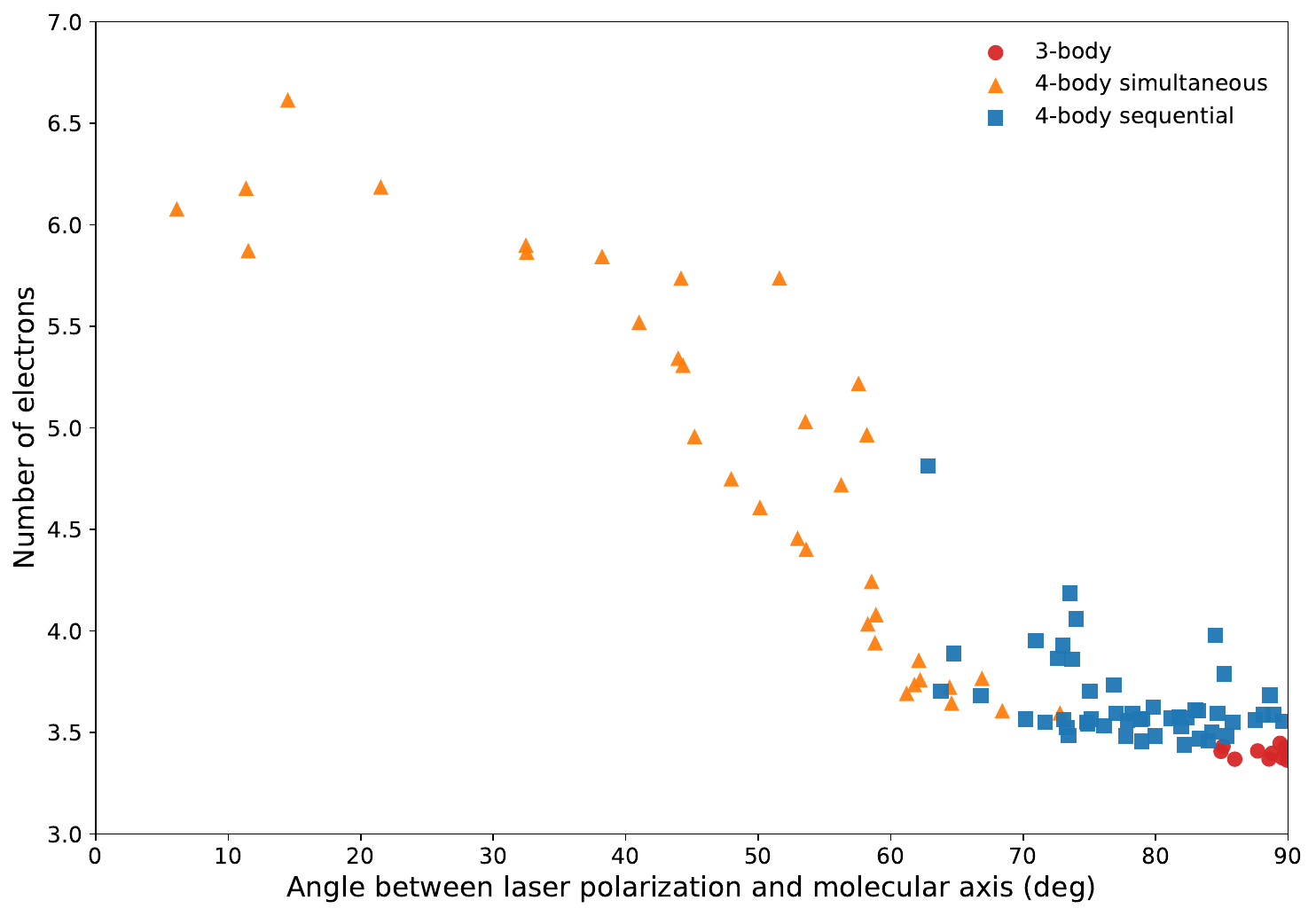}
  \caption{Scatter plot of the net ionization at $t=25$~fs
    versus the angle $\theta_{\mathrm{com}}$ between the laser polarization axis and the dimer molecular axis. Here $\theta_{\mathrm{com}}$ is defined as the angle between the polarization unit vector and the vector connecting the centers of mass of the two HCl monomers (the dimer COM axis). Points are colored by fragmentation outcome: 3-body events are colored as light blue; 4-body simultaneous events are colored as orange; 4-body sequential events are colored as dark blue. Angles are reported in $[0,90^\circ]$ because a linearly polarized field oscillates along its polarization axis, so $\hat{\mathbf{k}}$ and $-\hat{\mathbf{k}}$ are roughly equivalent.}
  \label{fig22}
\end{figure}

Fig.~\ref{fig22} reveals a pronounced orientation dependence of strong-field ionization: the net ionization increases strongly as the angle between the laser polarization and the dimer molecular axis decreases. Remarkably, the fragmentation outcomes are nearly partitioned by this single geometric parameter: trajectories with small alignment angles ($\theta_{\mathrm{com}}\approx 0$--$60^\circ$) exclusively yield near-simultaneous four-body breakup, whereas sequential four-body events predominantly populate larger angles and three-body outcomes are confined to the most misaligned configurations ($\theta_{\mathrm{com}}\gtrsim 80^\circ$). Together, these results identify laser--molecule alignment as a primary determinant of the ionization yield, which in turn predisposes the subsequent pathway selection. This is consistent with previous research on linear molecules.\cite{Russakoff2015PRA}

\begin{figure}[H]
  \centering
\includegraphics[width=0.5\textwidth]{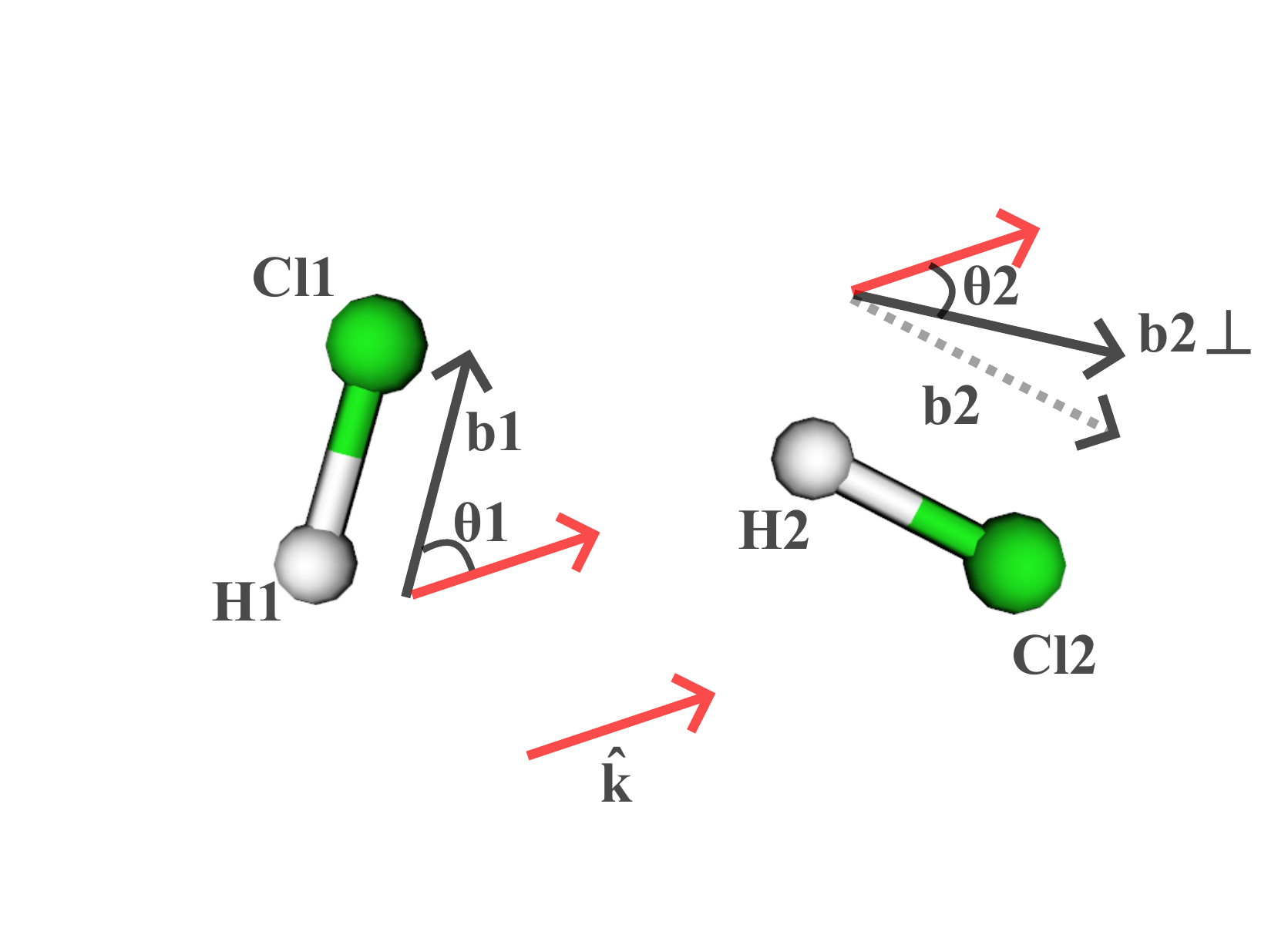}
\caption{Illustration of the two-angle orientation.
The horizontal axis is $\theta_1$, the angle between the laser polarization axis and the HCl1 
bond vector $\mathbf{b}_1$ (H1–Cl1). The vertical axis is $\theta_{2\perp}$, the angle between 
the polarization axis and the component of the HCl2 bond vector $\mathbf{b}_2$ (H2–Cl2) 
orthogonal to $\mathbf{b}_1$, i.e., 
$\mathbf{b}_{2\perp}=\mathbf{b}_2-(\mathbf{b}_2\!\cdot\!\hat{\mathbf{b}}_1)\hat{\mathbf{b}}_1$. }
  \label{fig24}
\end{figure}

To further characterize how molecular orientation controls ionization and channel branching, 
we introduce a bond-referenced two-angle description (Fig.~\ref{fig24}). 
The first angle, $\theta_1$, measures the alignment of the laser polarization axis 
with the HCl1 bond vector $\mathbf{b}_1$ (H1--Cl1) and therefore captures how strongly the 
field projects onto one monomer bond. The second angle, $\theta_{2\perp}$, is defined with 
respect to the orthogonalized direction of the second monomer bond, 
$\mathbf{b}_{2\perp}=\mathbf{b}_2-(\mathbf{b}_2\!\cdot\!\hat{\mathbf{b}}_1)\hat{\mathbf{b}}_1$, 
and is chosen to remove the trivial correlation that would arise if both angles were 
measured against the same axis. In this way, $(\theta_1,\theta_{2\perp})$ provides a compact 
measure of how strong the polarization projects onto the two HCl bonds in two independent geometric 
directions, which is the relevant condition for maximizing early-time ionization and promoting bond breaking.

\begin{figure}[H]
  \centering
\includegraphics[width=0.5\textwidth]{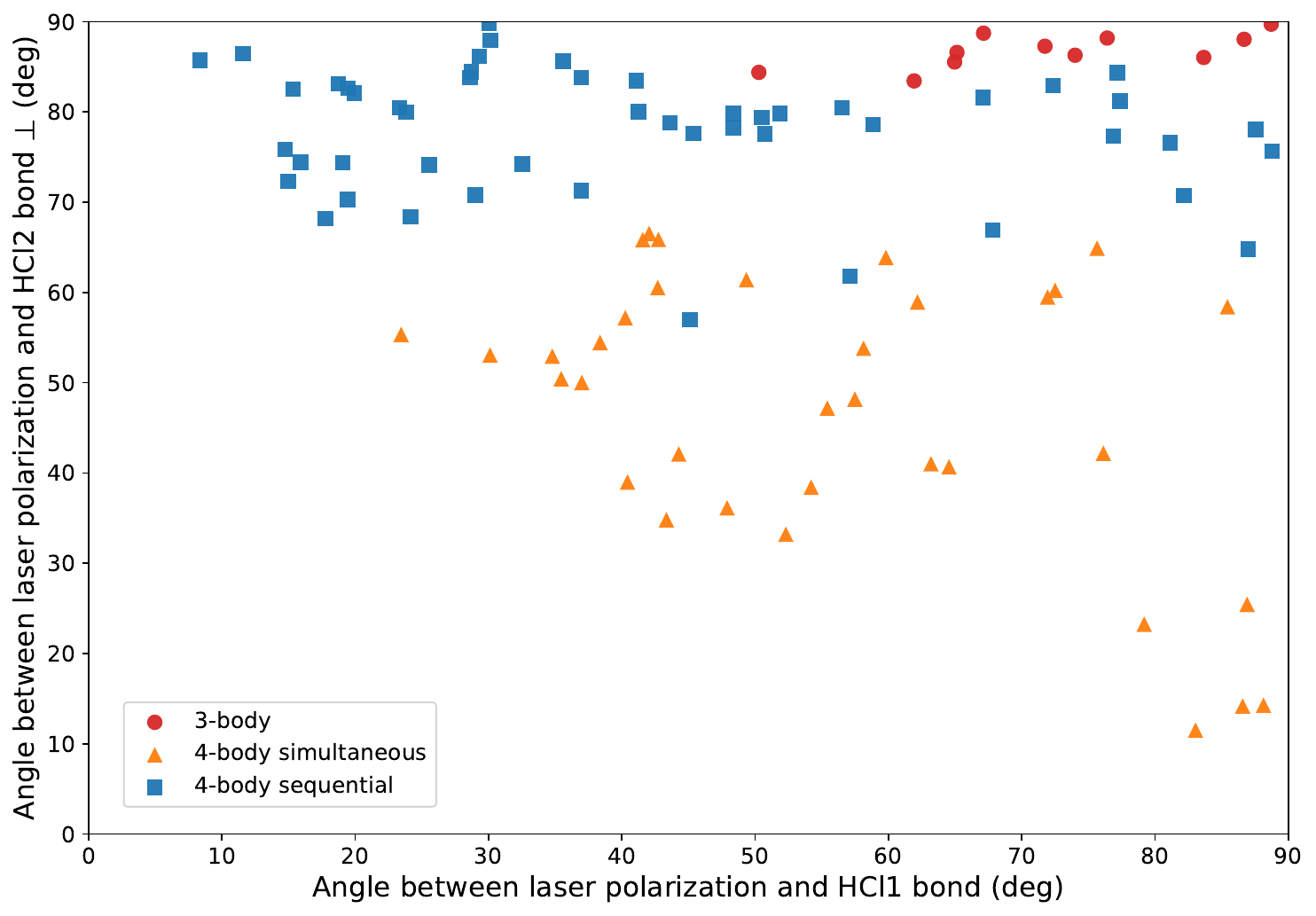}
  \caption{Two-angle orientation map for laser-driven ionization and channel 
  selectivity. Each point corresponds to one trajectory and is colored by fragmentation outcome. 
  Points are colored by fragmentation outcome: 3-body events are colored as light blue; 
  4-body simultaneous events are colored as orange; 4-body sequential events are colored as dark blue. 
  Angles are reported in $[0,90^\circ]$ because a linearly polarized field oscillates along 
  its polarization axis, so $\hat{\mathbf{k}}$ and $-\hat{\mathbf{k}}$ are roughly equivalent.}
  \label{fig23}
\end{figure}

The two-angle map 
in Fig.~\ref{fig23} resolves a clear channel selectivity 
in this coordinate frame. Three-body events cluster in a region where the polarization is 
poorly aligned with both directions: they occur predominantly at large $\theta_1$ 
(uniformly spread for $\theta_1\gtrsim 50^\circ$) together with 
large $\theta_{2\perp}$ ($\theta_{2\perp}\gtrsim 75^\circ$). This corner corresponds to configurations where the field has weak projections onto both bond-referenced axes, consistent with suppressed ionization and the survival of the HCl2 bond.

Sequential and simultaneous four-body events populate distinct $\theta_{2\perp}$ sectors. Near-simultaneous trajectories span a broad range of $\theta_1$ (approximately $20^\circ$--$90^\circ$) while concentrating at small-to-moderate $\theta_{2\perp}$ (roughly $0^\circ$--$60^\circ$), indicating that simultaneous two-bond dissociation is most often realized when the polarization has a relative substantial projection on the orthogonalized HCl2 direction. Sequential trajectories, by contrast, cover essentially the full $\theta_1$ range ($0^\circ$--$90^\circ$) but are preferentially concentrated at intermediate-to-large $\theta_{2\perp}$ (predominantly $\sim 55^\circ$--$80^\circ$, with the highest density near $\theta_{2\perp}\approx 70^\circ$), indicating a poorer alignment of the laser polarization with the HCl2 bond axis. This separation indicates that the relative alignment of the laser polarization with respect to the two H–Cl bond axes provides a systematic geometric bias: distinct orientation sectors preferentially drive the system toward either near-simultaneous or sequential two-bond breaking.

Taken together with the one-angle trend in Fig.~\ref{fig22}, these results connect molecular orientation to early-time ionization and provide a geometric interpretation of the channel branching. Strong alignment with the dimer axis increases the ionization and biases trajectories toward near-simultaneous four-body breakup, while large misalignment suppresses ionization and confines outcomes to sequential or three-body channels. The two-angle map further shows that this selectivity is not controlled by $\theta_1$ alone: the relative orientation with respect to the second, orthogonalized bond direction ($\theta_{2\perp}$) provides additional discrimination between the three pathways, revealing an extra bond-referenced geometrical constraint beyond simple alignment with the dimer axis.

\subsection{\label{sec:level6}Time-resolved dissociation dynamics}

\begin{figure}[H]
  \centering
\includegraphics[width=0.5\textwidth]{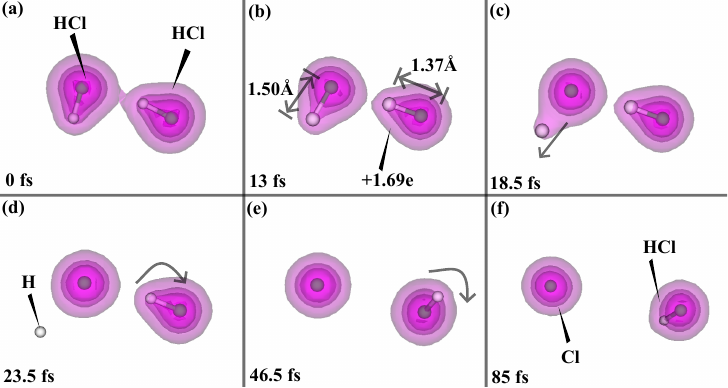}
  \caption{Representative snapshots along a three-body dissociation trajectory of the HCl dimer. At $t=13$~fs (panel~b), the net charge in the system is +3.05 e. The H1--Cl1 bond breaks irreversibly at $t=18.5$~fs (panel~c), whereas H2--Cl2 remains bound, yielding the three-body products \ce{H} + \ce{Cl} + \ce{HCl}. Panels~(a)--(f) are shown in the same view. Four electron-density isosurfaces are displayed in each panel. The isosurface values shown are 0.10, 0.567, 1.033, and 1.50.}
  \label{fig8}
\end{figure}

To connect these statistical trends to real-space dynamics, we next present representative trajectory snapshots for each channel together with time-resolved traces of (i) the laser electric field and the remaining electron number, and (ii) the laser electric field overlaid with the H--Cl bond-length evolution.
These time-domain diagnostics provide an intuitive picture of when charge is built up relative to the onset of bond breaking and how the nuclear motion diverges among the three channels.

\begin{figure}[H]
  \centering
\includegraphics[width=0.5\textwidth]{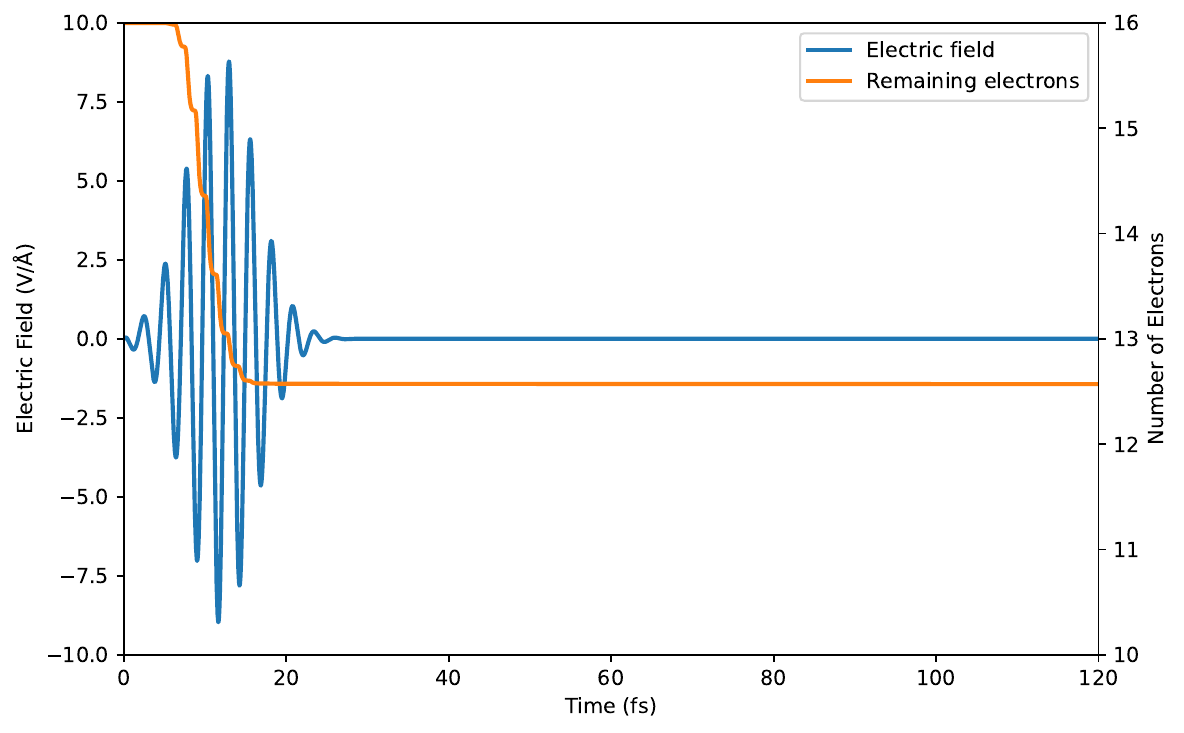}
  \caption{Time evolution of the number of electrons remaining in the simulation box for the representative trajectory shown above, overlaid with the driving laser electric field. The total ionization for this run is $3.43~e$.}
  \label{fig9}
\end{figure}

\begin{figure}[H]
  \centering
\includegraphics[width=0.5\textwidth]{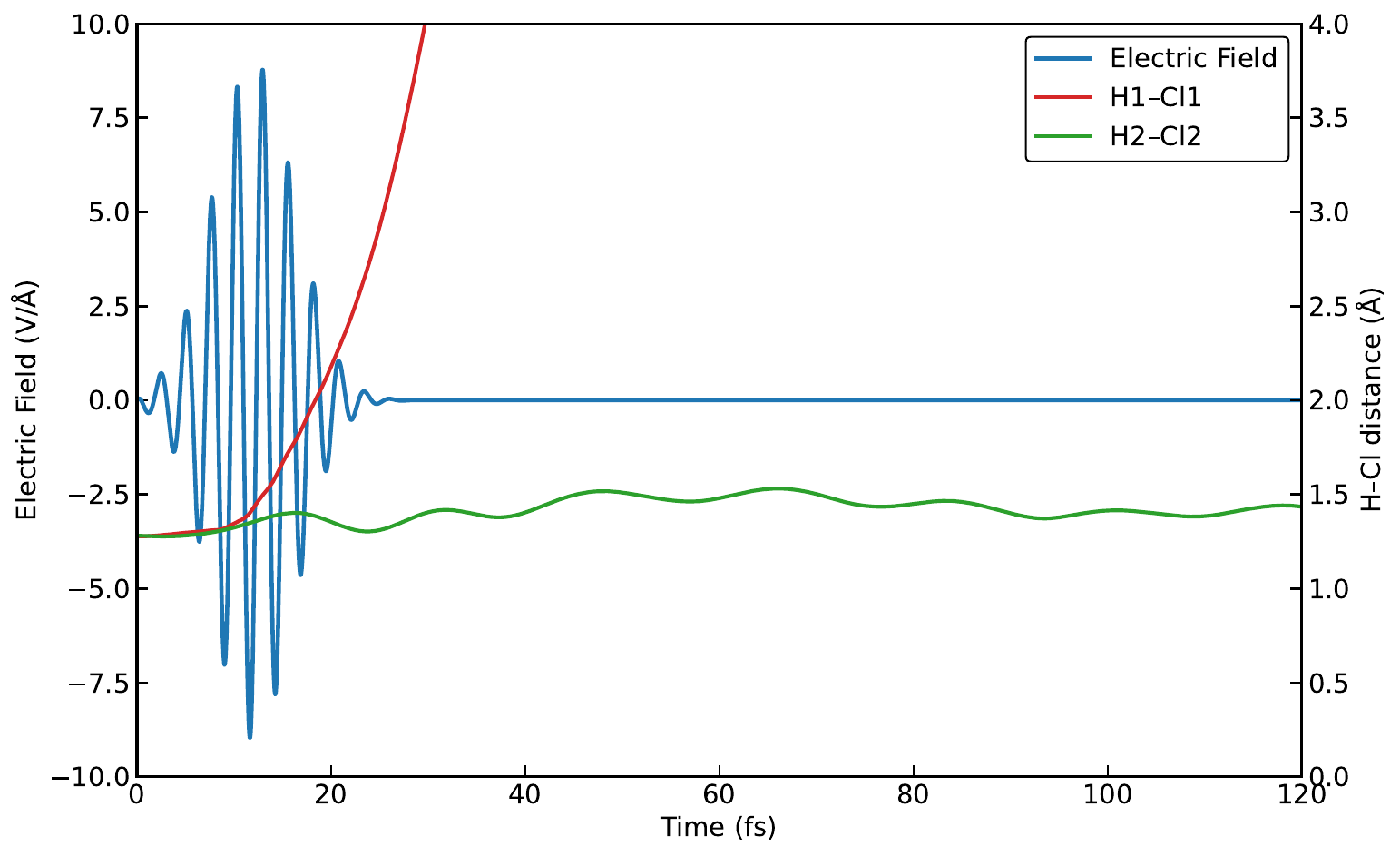}
  \caption{Time evolution of the two H--Cl internuclear distances for the representative trajectory shown above, overlaid with the driving laser electric field. The initial internuclear distance is 1.274\AA.}
  \label{fig10}
\end{figure}

Figs.~\ref{fig8}, \ref{fig9}, and \ref{fig10} show, respectively, representative snapshots, the time-dependent remaining electron number overlaid with the driving laser electric field, and the H--Cl internuclear distances overlaid with the same field, for a representative three-body dissociation trajectory. From Fig.~\ref{fig9}, noticeable ionization onsets at $\sim 7$~fs when the electric field reaches 
$\sim 5~\mathrm{V}$/\AA. The ionization proceeds predominantly during the rising and 
near-peak portion of the pulse and largely saturates by $\sim 15$~fs. Importantly, this 
saturation occurs even though the electric field remains relatively strong (still 
above $\sim 6~\mathrm{V}$/\AA), suggesting that the cessation of ionization is governed by electronic depletion and/or charge-state saturation rather than by the instantaneous electric field amplitude alone. A total of 3.43 electrons are ionized in this trajectory.

Between $t=0$~fs [Fig.~\ref{fig8}(a)] and $t=13$~fs [Fig.~\ref{fig8}(b)], both H--Cl bonds elongate from the initial $1.27$~\AA, with H1--Cl1 stretching more noticeably to $1.50$~\AA, while H2--Cl2 reaches $1.37$~\AA ~at 13~fs. At this time, the charge localized on HCl2 is $+1.69~e$ out of a net charge of $+3.05~e$. At $t=18.5$~fs [Fig.~\ref{fig8}(c)], H1--Cl1 meets our bond-breaking criterion; the ejected H1 fragment becomes clearly separated from its parent molecule by $t=23.5$~fs [Fig.~\ref{fig8}(d)]. Notably, the emission direction of H1 deviates only modestly from the initial H1--Cl1--H2 plane/angle. Owing to the combined Coulomb repulsion of the departing H1 and the residual Cl1, a pronounced rotation of HCl2 about Cl2 is initiated at $\sim 25$~fs, as illustrated by the structural evolution in Fig.~\ref{fig8}(d)--(f), while the HCl2 fragment also begins to recoil away from Cl1. From Fig.~\ref{fig10}, H2--Cl2 undergoes substantial vibrational motion yet remains bound throughout the full simulation window, consistent with the three-body outcome \ce{H} + \ce{Cl} + \ce{HCl}.

\begin{figure}[H]
  \centering
\includegraphics[width=0.5\textwidth]{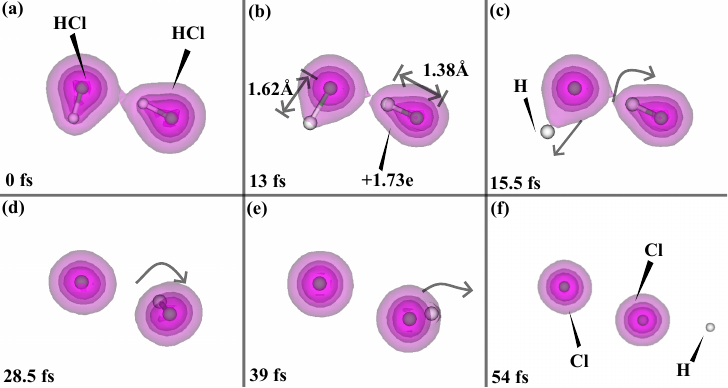}
  \caption{Representative snapshots along a four-body sequential dissociation trajectory of the HCl dimer. At $t=13$~fs (panel~b), the net charge in the system is +3.13 e. The H1--Cl1 bond breaks irreversibly at $t=15.5$~fs (panel~c), whereas H2--Cl2 remains bound until it breaks at $t=39$~fs (panel~e), yielding the four-body products \ce{H} + \ce{Cl} + \ce{H} + \ce{Cl}. Panels~(a)--(f) are shown in the same view. Four electron-density isosurfaces are displayed in each panel. The isosurface values shown are 0.10, 0.567, 1.033, and 1.50.}
  \label{fig11}
\end{figure}

\begin{figure}[H]
  \centering
\includegraphics[width=0.5\textwidth]{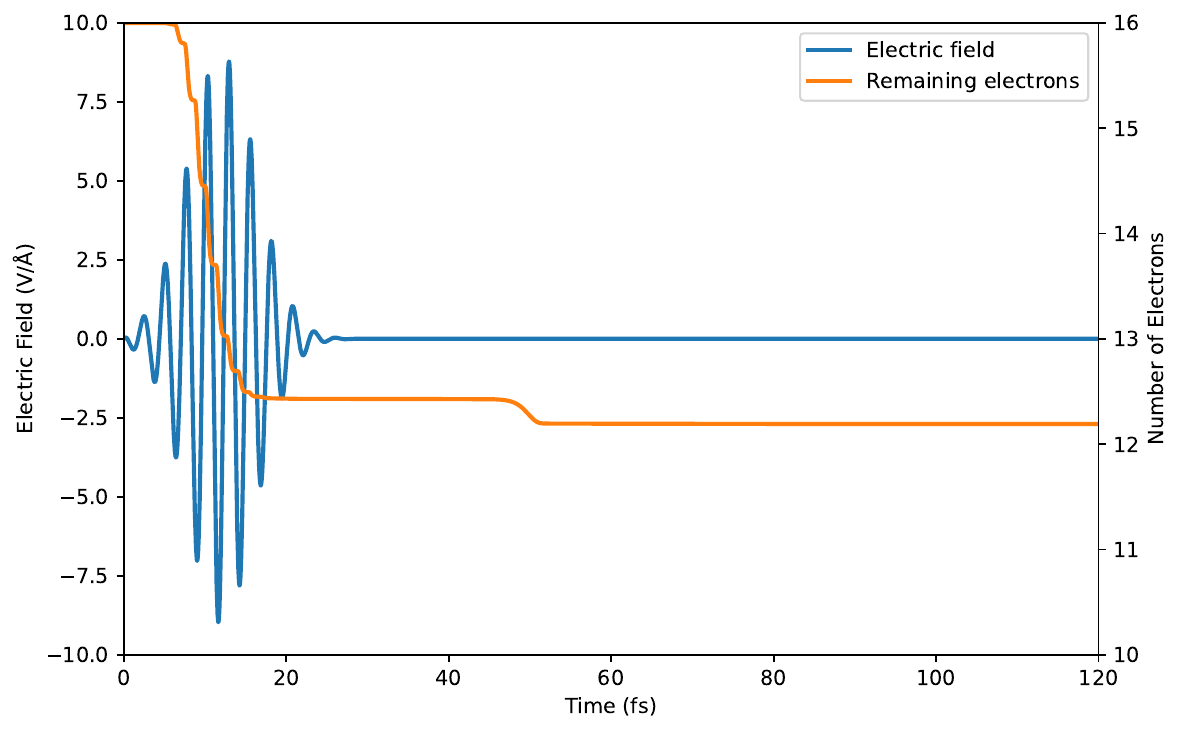}
  \caption{Time evolution of the number of electrons remaining in the simulation box for the representative trajectory shown above, overlaid with the driving laser electric field. The total ionization for this run is $3.57~e$. The additional decrease in the number of electrons at around 50fs is because the outgoing electronic density associated with the early-emitted H1 fragment reaches the complex absorbing boundary (CAP) and is removed from the simulation box}
  \label{fig12}
\end{figure}

\begin{figure}[H]
  \centering
\includegraphics[width=0.5\textwidth]{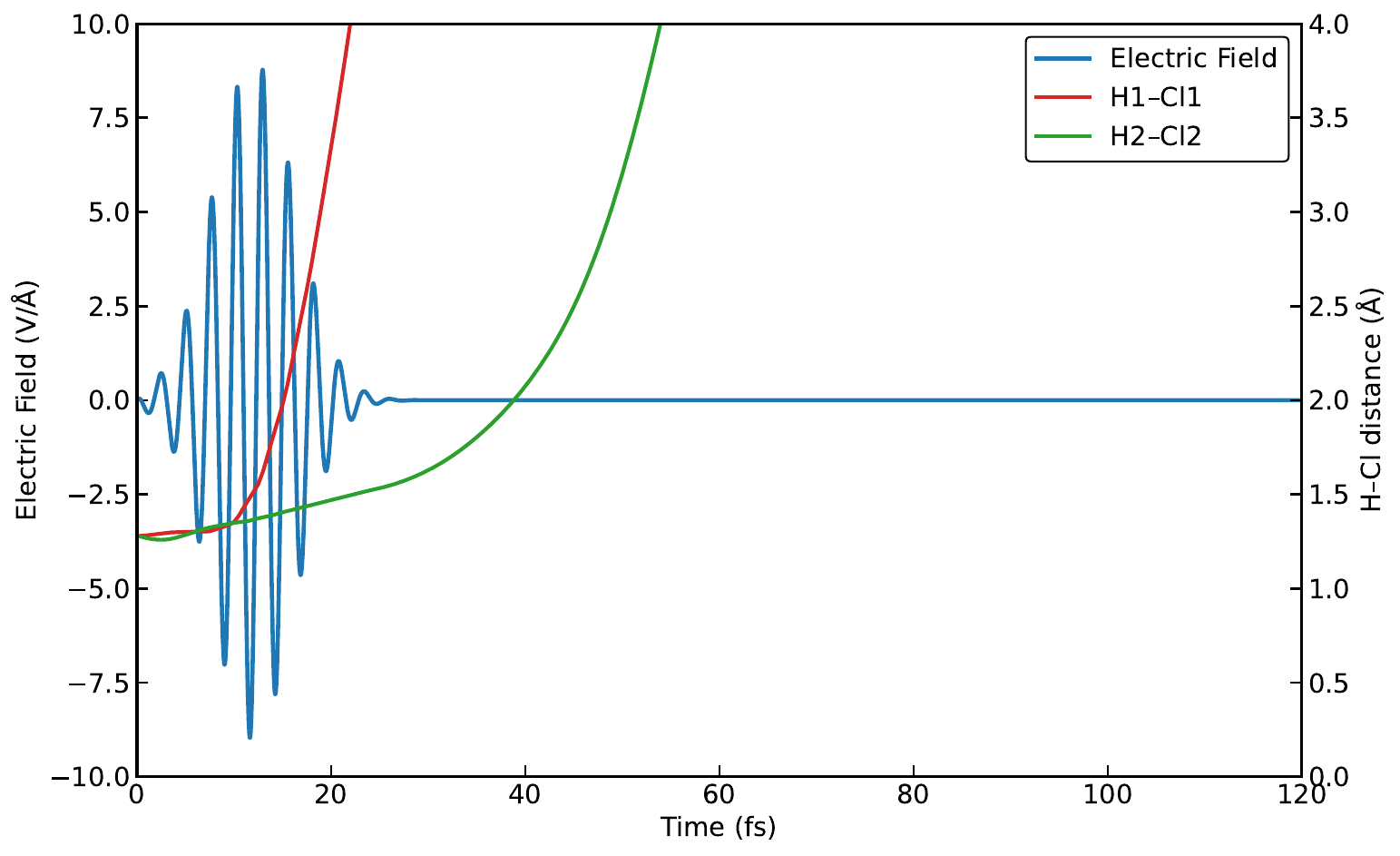}
  \caption{Time evolution of the two H--Cl internuclear distances for the representative trajectory shown above, overlaid with the driving laser electric field. The initial internuclear distance is 1.27\AA.}
  \label{fig13}
\end{figure}

Figs.~\ref{fig11}--\ref{fig13} present a representative four-body sequential dissociation trajectory, shown in the same format as Figs.~\ref{fig8}--\ref{fig10}. 
The ionization timing relative to the pulse electric field is comparable to the three-body case, but the overall total ionization is larger, reaching $3.57~e$ for this trajectory. 
At the early-time snapshot $t=13$~fs (Fig.~\ref{fig11}b), both H--Cl bonds are already elongated, with H1-Cl1 distance 1.62 \AA~ and H2-Cl2 distance 1.38 \AA. 
At the same time, the HCl2 molecule carries a substantial fraction of charge, with 1.73 e out of a net charge of $+3.13~e$.

The first bond breaks early in this trajectory: H1--Cl1 meets the bond-breaking criterion at $t=15.5$~fs (Fig.~\ref{fig11}c). 
The breaking of the second bond is delayed until $t=39$~fs (Fig.~\ref{fig11}e), corresponding to an inter-break time of $\Delta t = 23.5$~fs, which places this event unambiguously in the sequential regime.
Immediately following the first bond dissociation, a pronounced rotational motion of HCl2 about Cl2 is initiated; this rotation is clearly visible across Fig.~\ref{fig11}c--e. 
Because the first bond dissociation happens earlier here, the onset of this rotation is correspondingly earlier than in the three-body representative trajectory.

The delayed second bond dissociation is captured by the H2--Cl2 distance trace (Fig.~\ref{fig13}). 
From $t=0$ to $\sim 30$~fs, the internuclear distance between H2 and Cl2 increases in an approximately linear fashion, indicating a gradual destabilization rather than bounded oscillation. Beyond $\sim 30$~fs (when H2 and Cl2 separate by $\approx 1.5~\text{\AA}$), 
the stretching accelerates, and H2--Cl2 reaches the dissociation threshold at $t = 39$~fs (Fig.~\ref{fig11}e), completing the sequential four-body breakup.

\begin{figure}[H]
  \centering
\includegraphics[width=0.5\textwidth]{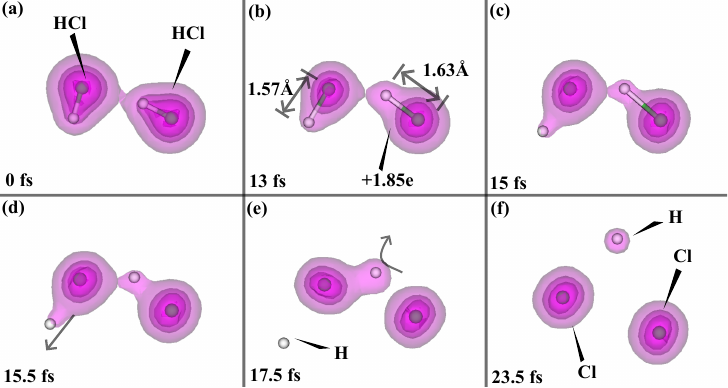}
  \caption{Representative snapshots along a four-body simultaneous dissociation trajectory of the HCl dimer. At $t=13$~fs (panel~b), the net charge in the system is +4.09 e. The H1--Cl1 bond breaks irreversibly at $t=15$~fs (panel~c), whereas H2--Cl2 breaks almost simultaneously at $t=15.5$~fs (panel~d), yielding the four-body products \ce{H} + \ce{Cl} + \ce{H} + \ce{Cl}. Panels~(a)--(f) are shown in the same view. Four electron-density isosurfaces are displayed in each panel. The isosurface values shown are 0.10, 0.567, 1.033, and 1.50.}
  \label{fig14}
\end{figure}

\begin{figure}[H]
  \centering
\includegraphics[width=0.5\textwidth]{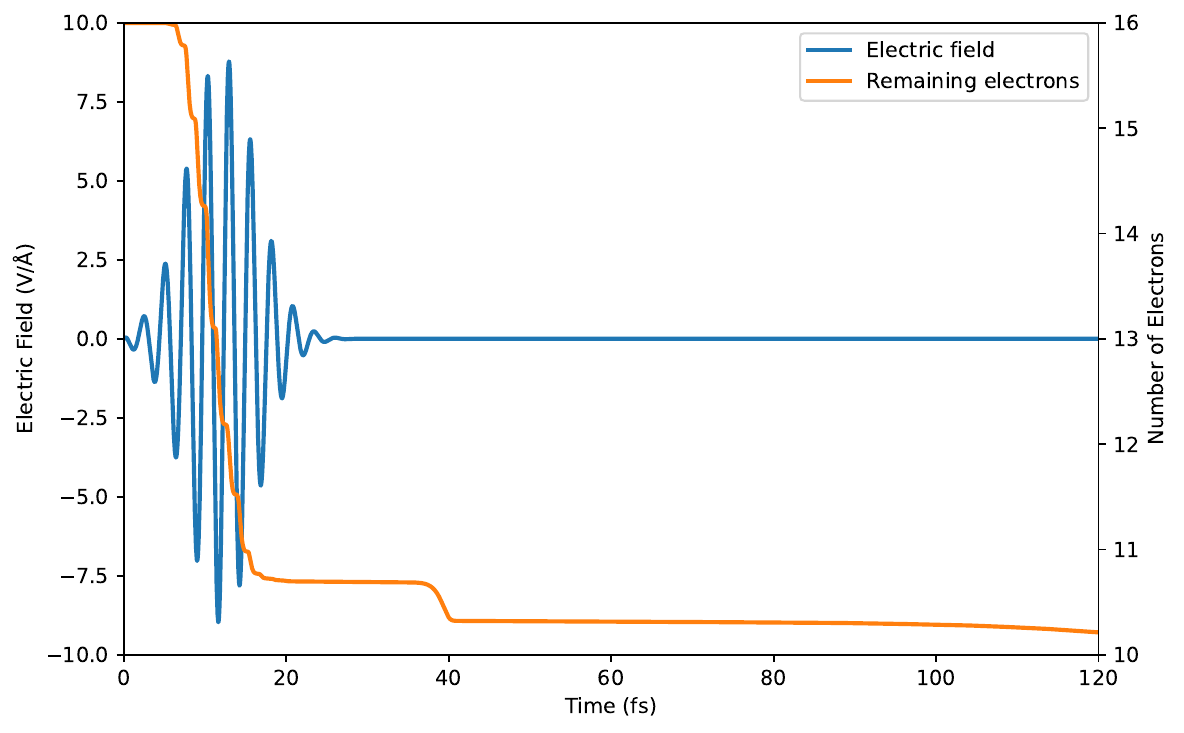}
  \caption{Time evolution of the number of electrons remaining in the simulation box for the representative trajectory shown above, overlaid with the driving laser electric field. The total ionization for this run is $5.31~e$. The additional decrease in the number of electrons at around 40fs is because the outgoing electronic density associated with the emitted H2 fragment reaches the complex absorbing boundary (CAP) and is removed from the simulation box}
  \label{fig15}
\end{figure}

\begin{figure}[H]
  \centering
\includegraphics[width=0.5\textwidth]{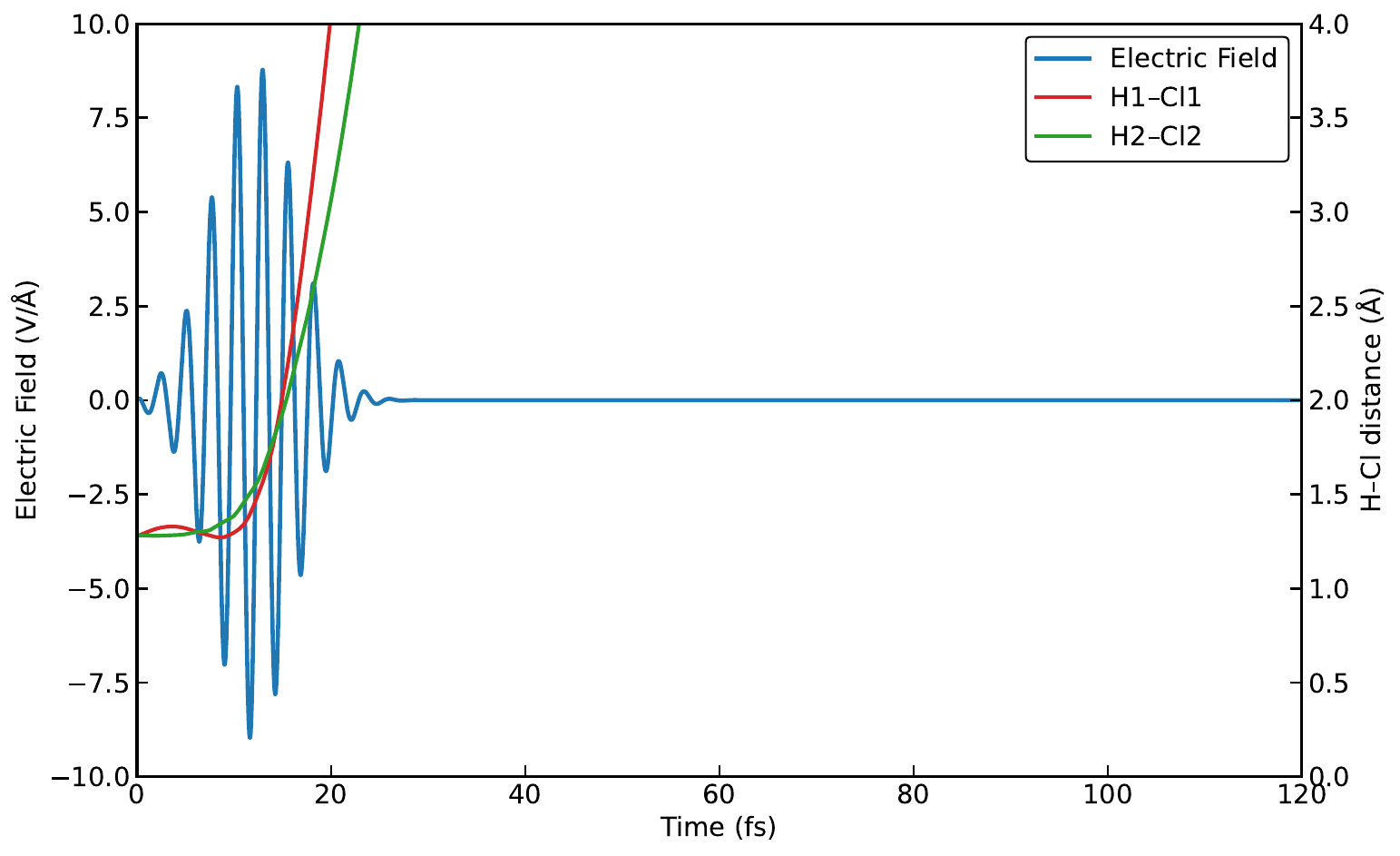}
  \caption{Time evolution of the two H--Cl internuclear distances for the representative trajectory shown above, overlaid with the driving laser electric field. The initial internuclear distance is 1.27\AA.}
  \label{fig16}
\end{figure}

Figs.~\ref{fig14}--\ref{fig16} present a representative four-body simultaneous dissociation trajectory, shown in the same format as Figs.~\ref{fig8}--\ref{fig10}. 
The ionization timing relative to the pulse electric field is similar to the other channels (Fig.~\ref{fig15}), but this trajectory exhibits substantially stronger ionization, reaching a total ionization of $5.31~e$. 
At the early-time snapshot $t=13$~fs (Fig.~\ref{fig14}b), both H--Cl bonds are already strongly elongated, with H1-Cl1 distance 1.57~\AA\ and H2-Cl2 distance 1.63~\AA. 
Consistent with the ensemble-level trends, both the total ionization and the HCl2 fragment charge are the highest among the three representative trajectories at this time: the system carries a net charge of $+4.09~e$, of which HCl2 accounts for 1.85~e.

In this trajectory, the two bonds dissociate occur within a narrow time window. 
H1--Cl1 meets the bond-breaking criterion at $t=15.0$~fs (Fig.~\ref{fig14}c), and the H2--Cl2 breaks shortly thereafter (Fig.~\ref{fig14}d), yielding an inter-break delay that falls in the simultaneous regime. 
Following the second dissociation, the H2 fragment is emitted initially toward Cl1; after reaching a closest separation of $\sim 2.1$~\AA(Fig.~\ref{fig14}e), it is deflected and subsequently moves away as the fragments separate(Fig.~\ref{fig14}e-f).

The rapid two-bond breakup is also evident in the internuclear-distance traces (Fig.~\ref{fig16}). 
For H1--Cl1, the distance shows a non-monotonic early-time evolution: it first increases, then decreases, and finally increases again, producing a clear inflection. 
This behavior reflects the fact that the randomly sampled initial nuclear motion in this particular trajectory initially stretches the H1--Cl1 bond before the laser-driven dynamics dominates. By contrast, the distance between H2 and Cl2 grows more gradually at early times, consistent with H2 being transiently ``caged'' between the two chlorine atoms, and then accelerates as the Coulomb-driven separation strengthens. 
Overall, Fig.~\ref{fig16} shows that both H--Cl distances reach the $2.0$~\AA\ dissociation threshold within nearly the same time window, consistent with the simultaneous four-body breakup mechanism.

\begin{figure}[H]
  \centering
\includegraphics[width=0.5\textwidth]{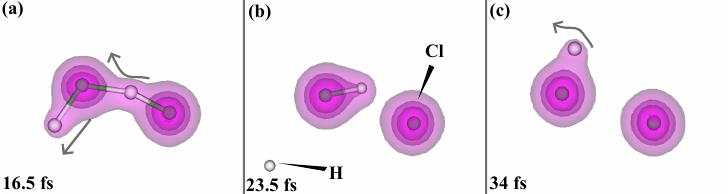}
  \caption{Representative snapshots along a four-body simultaneous dissociation trajectory of the HCl dimer in which a temporary bond is formed between H2 and Cl1. Panels~(a)--(c) are shown in the same view. Four electron-density isosurfaces are displayed in each panel. The isosurface values shown are 0.10, 0.567, 1.033, and 1.50.}
  \label{fig17}
\end{figure}

In a small subset of trajectories(4/96), we observe a transient short-range contact between H2 and Cl1,
i.e., a temporary ``bond-switching'' motif in which H2 approaches Cl1 and forms a short-lived H2--Cl1 interaction
before separating again. Representative snapshots are shown in Fig. \ref{fig17}.
These events do not constitute a distinct final fragmentation channel as the asymptotic products still fall into the three classes defined above,
and are therefore not treated as a separate pathway in the branching statistics.

We note that both the four-body sequential and simultaneous pathways can exhibit a broad distribution of the H--H emission angle.
In both cases, the emission direction of H1 is comparatively more constrained, remaining closer to its local dissociation axis at the moment of the first bond dissociation, whereas the H2 emission direction is substantially more variable and therefore dominates the spread of $\theta_{\mathrm{HH}}$.

For the sequential channel, the second H--Cl bond dissociation occurs while HCl2 undergoes pronounced rotational motion about Cl2.
As a result, the instantaneous orientation of the H2--Cl2 bond axis at the breaking time can differ markedly from event to event, so sampling different initial conditions naturally leads to a wide range of $\theta_{\mathrm{HH}}$.

For the simultaneous channel, H2 is initially accelerated toward the vicinity of Cl1 and subsequently experiences strong trajectory bending in the multi-center Coulomb field.
This scattering-like deflection makes the final H2 emission direction highly sensitive to the ensemble of initial conditions, including the thermally sampled initial nuclear velocities and the randomized laser polarization, again producing large variations in $\theta_{\mathrm{HH}}$.

Taken together, the three representative trajectories provide a real-space and time-resolved interpretation of the charge-dependent trends established.
The three-body case is characterized by a lower early-time ionization and a bound, rotating HCl2 fragment after the first dissociation.
The sequential four-body case exhibits a modestly higher charge on HCl2 and a delayed, gradually destabilized H2--Cl2 bond that accelerates into an unbound state only after substantial post-ionization nuclear evolution.
In contrast, the simultaneous four-body case corresponds to the highest early-time ionization, leading to two H--Cl bond dissociations within a narrow time window and strong multi-center Coulomb deflection that sensitizes the final H2 emission direction.

These mechanistic differences are expected to leave clear signatures in fragmentation observables measured at long times, including the kinetic-energy release (KER) and the angular correlations between the fragments.
In the next section, we therefore present the channel-resolved distributions of KER, $\theta_{\mathrm{HH}}$, and $\theta_{\mathrm{ClCl}}$, and compare them with the available experimental data.

\subsection{\label{sec:level7}KER and angular distributions}

For direct comparison with the experimentally reported four-body breakup\cite{Zhao2025HClDimer}, we focus on the 4-body KER distributions in the main text.
The total KER distribution including the minority 3-body channel is provided in the Supporting Information.

\begin{figure}[H]
  \centering
\includegraphics[width=0.5\textwidth]{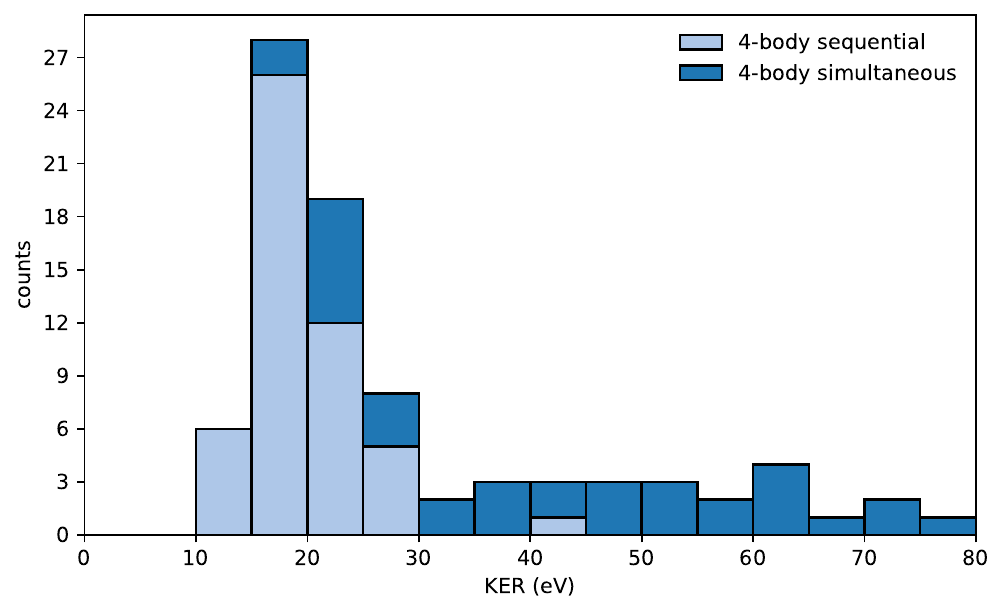}
  \caption{Total KER distribution for all four-body dissociation events (in eV). Histograms are shown as counts with a bin width of 5~eV and stacked to indicate the separate contributions from the two four-body mechanisms: the sequential channel (light blue) and the simultaneous channel (dark blue).}
  \label{fig18}
\end{figure}

Fig.~\ref{fig18} shows the total KER distribution for all four-body events, plotted as a stacked histogram to separate the contributions from the sequential and simultaneous pathways.
The spectrum spans a broad range from $\sim 10$ to $\sim 80$~eV and is strongly right-skewed (positively skewed): most events populate the low-energy region while a weak, long high-energy tail extends to large KER.
The dominant maximum occurs in the 15--20~eV bin, with a secondary maximum in the 20--25~eV bin. Channel-resolved contributions reveal that the low-energy part (roughly 10--30~eV) is dominated by the sequential pathway, with a smaller but non-negligible simultaneous contribution, whereas the high-energy tail above $\sim 30$~eV arises almost exclusively from simultaneous breakup events.

Quantitatively, the KER distribution of all four-body events has a mean of
$\langle \mathrm{KER}\rangle = 29.30$~eV with a median of 21.63~eV and a standard deviation of 16.30~eV,
consistent with the pronounced right-skewed profile (mean $\gg$ median).
Channel-resolved statistics further highlight the bimodal energetic character:
sequential events cluster at low energies with
$\langle \mathrm{KER}\rangle = 20.07$~eV (median 19.37~eV, $\sigma=4.86$~eV),
whereas simultaneous events are substantially more energetic and heterogeneous with
$\langle \mathrm{KER}\rangle = 42.49$~eV (median 40.81~eV, $\sigma=17.83$~eV).

\begin{figure}[H]
  \centering
\includegraphics[width=0.5\textwidth]{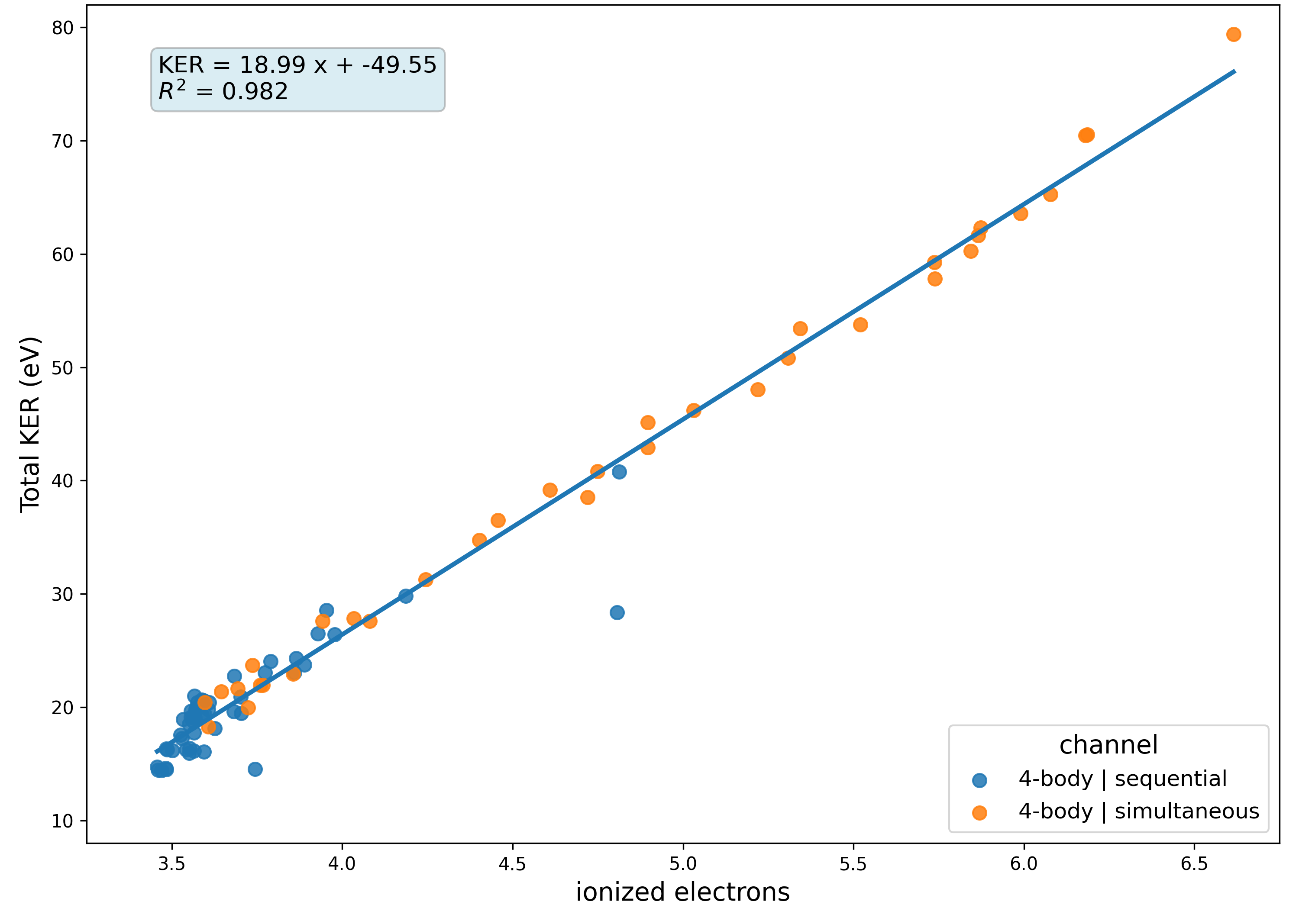}
  \caption{Scatter plot of the total kinetic energy release (KER)
  versus the total ionization(at 25 fs, when the laser field has essentially vanished) by the laser for four-body dissociation events. A linear regression yields a strong correlation ($R^{2}=0.982$), $\mathrm{KER} = 18.99\,q - 49.55$, where $q$ is the total ionization and KER is in eV. Points are colored by channel: sequential (blue) and simultaneous (orange).}
  \label{fig19}
\end{figure}

Fig.~\ref{fig19} correlates the total kinetic energy release (KER) with the total ionization $q$ for the four-body dissociation events. 
A clear, near-linear trend is observed: trajectories that ionize more strongly systematically yield larger KER. 
A least-squares fit gives $\mathrm{KER}=18.99\,q-49.55$ (eV) with $R^{2}=0.982$, indicating that $q$ captures the dominant variation in the final energy release within the four-body ensemble.

This behavior is physically expected because, once the system reaches a given charge state, the subsequent breakup is largely governed by Coulomb repulsion between the emerging fragments.
Higher total ionization increases the effective Coulomb interaction and thus the amount of potential energy converted into fragment kinetic energy, leading to a higher KER. Consistent with the charge-resolved channel statistics, the simultaneous channel (orange) populates the higher-$q$ / higher-KER region more frequently, whereas the sequential channel (blue) is concentrated at lower ionization and correspondingly lower KER.

With the overall KER distribution and its strong charge dependence established, we now compare the simulated spectrum with the experimental measurements. 

Zhao \textit{et al.} reported a four-body KER spectrum for the detected channel $\mathrm{H^+ + H^+ + Cl^+ + Cl^+}$ that is concentrated primarily between 10 and 40~eV with a prominent peak around 24~eV.\cite{Zhao2025HClDimer}
In our full four-body ensemble, the main weight is shifted slightly to lower energies (peaks in the 15--25~eV range), which is consistent with the fact that sequential events dominate our statistics (50/85) and are associated with a comparatively lower mean ionization (3.66$\,e$) than the quadruple ionization in the experiment.
This interpretation is quantitatively supported by the strong correlation between KER and the total ionization (Fig.~\ref{fig19}).
Evaluating the linear regression line on Fig. \ref{fig19} at $N_{\mathrm{ion}}=4$ gives $\mathrm{KER}\approx 26.4$~eV, close to the experimental peak energy, suggesting that the remaining discrepancy is largely attributable to charge-state selection.
Indeed, when we restrict the analysis to higher-ionization trajectories (e.g., $N_{\mathrm{ion}}>3.8$; see Supporting Information), the peak shifts to the higher energy range.
Finally, our spectrum is visibly broader and extends to $\sim 80$~eV because a subset of simultaneous events reaches much higher ionization (up to $\gtrsim 6.6\,e$), producing a long high-energy tail; such highly charged breakups would be effectively excluded in the experiment by ion/charge-state selectivity. 
Residual differences are also expected because we can only model the
experimental laser shape and slight differences
can change the temporal ionization profile and the relative population of charge states.

\begin{figure}[H]
  \centering
\includegraphics[width=0.5\textwidth]{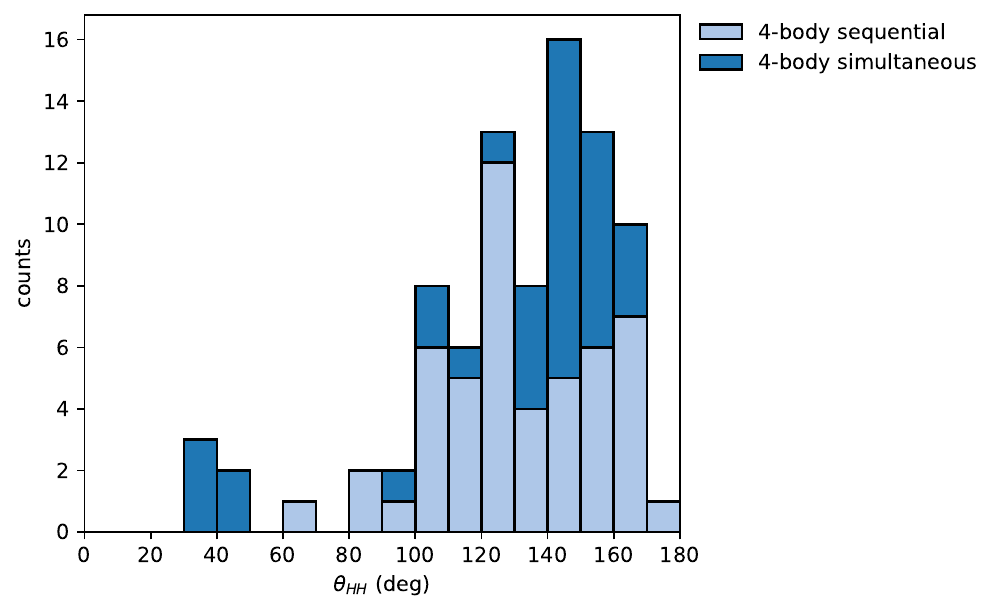}
  \caption{$\theta_{\mathrm{HH}}$ distribution for all four-body dissociation events. Histograms are shown as counts with a bin width of 10 degree and stacked to indicate the separate contributions from the two four-body mechanisms: the sequential channel (light blue) and the simultaneous channel (dark blue).}
  \label{fig20}
\end{figure}

We next examine the momentum-space emission-angle distributions and compare them with the experimental trends reported by Zhao \textit{et al.}~\cite{Zhao2025HClDimer}.

\begin{figure}[H]
  \centering
\includegraphics[width=0.5\textwidth]{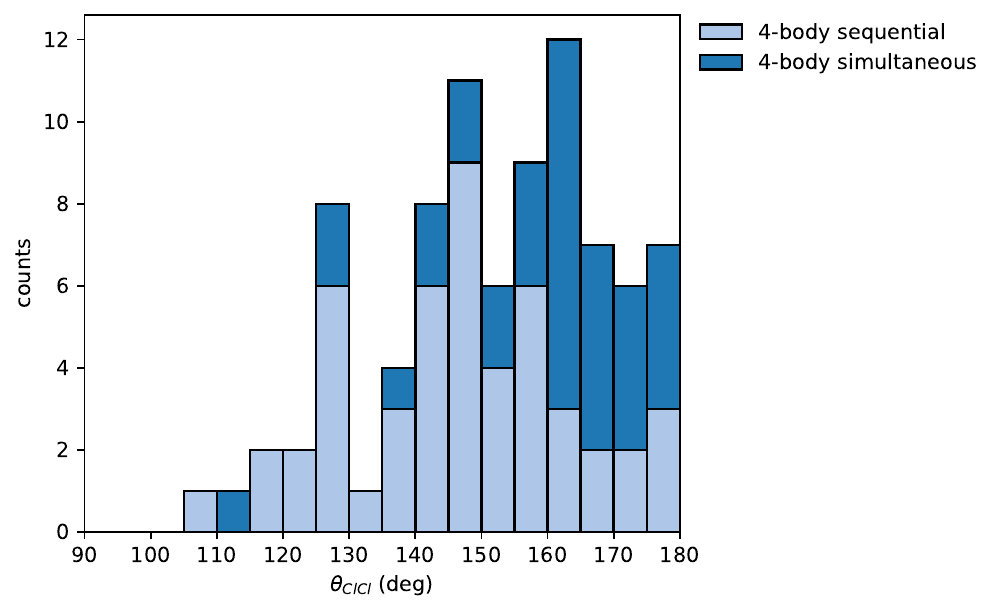}
  \caption{$\theta_{\mathrm{ClCl}}$ distribution for all four-body dissociation events. Histograms are shown as counts with a bin width of 5 degree and stacked to indicate the separate contributions from the two four-body mechanisms: the sequential channel (light blue) and the simultaneous channel (dark blue).}
  \label{fig21}
\end{figure}

The H--H emission angle $\theta_{\mathrm{HH}}$ spans a broad range from $\sim 30^\circ$ to $180^\circ$, with the dominant population concentrated in the $\sim 120^\circ$--$170^\circ$ region and a maximum in the $140^\circ$--$150^\circ$ bin (Fig.~\ref{fig20}). 
Across most of the high-probability region ($\sim 100^\circ$--$170^\circ$), both channels contribute, with a noticeable enhancement of the simultaneous pathway near the main peak ($\sim 140^\circ$--$160^\circ$) and an exclusive low-angle tail ($\sim 30^\circ$--$50^\circ$) for the simultaneous channel. 
We note, however, that this low-angle region contains only a few events and is therefore discussed as a qualitative tendency rather than a statistically robust selectivity. 
Overall, the broad $\theta_{\mathrm{HH}}$ distribution is qualitatively consistent with Zhao \textit{et al.}~\cite{Zhao2025HClDimer}, who also report a wide H--H angular spread with the highest population occurring between $\sim 90^\circ$ and $\sim 170^\circ$. In contrast, the Cl--Cl emission angle $\theta_{\mathrm{ClCl}}$ is confined to a narrower interval, ranging from $\sim 105^\circ$ to $180^\circ$, with most events lying in the $140^\circ$--$180^\circ$ sector (Fig.~\ref{fig21}). 
The distribution exhibits a primary peak at $160^\circ$--$165^\circ$ and a secondary maximum at $145^\circ$--$150^\circ$. 
Channel-resolved statistics show that the simultaneous pathway contributes disproportionately in the near-back-to-back region ($160^\circ$--$180^\circ$), whereas the sequential pathway dominates the remaining angular bins. 
These trends are again in qualitative agreement with the experiment, where $\theta_{\mathrm{ClCl}}$ is substantially more confined than $\theta_{\mathrm{HH}}$ and is primarily distributed in the $\sim 130^\circ$--$180^\circ$ range.
A detailed mechanistic attribution of the channel-dependent anisotropy would likely require correlating the emission angles with the instantaneous geometry and charge partitioning at the time of bond breaking, which we leave for future work.

Taken together, the KER and emission-angle observables provide an independent, experimentally accessible validation of the charge-dependent picture established above. 
Within the four-body ensemble, the strong (near-linear) KER--$q$ correlation indicates that early-time ionization largely sets the available Coulomb energy scale, while the channel-resolved histograms show that sequential events predominantly populate the low-energy portion of the spectrum and simultaneous events are responsible for the high-energy tail. 
The angular distributions further demonstrate a clear hierarchy of anisotropy, with $\theta_{\mathrm{HH}}$ broadly distributed and $\theta_{\mathrm{ClCl}}$ substantially more confined, in qualitative agreement with the experimental trends. 
Overall, these results support a coherent mechanistic interpretation in which early-time ionization predisposes the subsequent breakup pathway and imprints itself on both energetic (KER) and kinematic (emission-angle) fragmentation signatures.

\section{Conclusion}

We have presented a channel-resolved, time-resolved study of laser-driven Coulomb explosion of the HCl dimer using real-time TDDFT on a real-space grid with explicit inclusion of the driving laser field. By analyzing an ensemble of 96 trajectories with randomized polarization directions and thermally sampled initial nuclear velocities, we identify three dominant fragmentation pathways: three-body breakup, four-body sequential breakup, and four-body simultaneous breakup. In all trajectories, H1--Cl1 dissociates first within a narrow time window, after which the system either retains a bound HCl fragment (three-body) or undergoes a second H--Cl dissociation. For four-body events, an inter-break delay $\Delta t$ provides a practical discriminator between simultaneous and sequential mechanisms.

A central result is that pathway selection is strongly correlated with early-time ionization during the laser interaction. The total ionization at $t=13$~fs already provides clear separation between the two four-body mechanisms, with simultaneous events exhibiting higher and more heterogeneous ionization than sequential events. Moreover, although three-body and sequential events have similar total ionization, the HCl2 fragment charge shows a small but systematic upward shift for sequential trajectories, indicating that the additional ionization distinguishing this pathway is preferentially localized on the HCl2 molecule. Complementarily, the laser–molecule orientation analysis shows that alignment with the dimer and bond axes strongly modulates the early-time ionization and thereby biases the channel branching.

Long-time, experimentally accessible observables reflect these charge-dependent trends. The four-body KER distribution is right-skewed and is dominated at low energies by sequential breakup, while the high-energy tail arises primarily from simultaneous events. The KER correlates strongly with the total ionization, consistent with Coulombic conversion of electrostatic potential energy into fragment kinetic energy. The emission-angle distributions show a broad $\theta_{\mathrm{HH}}$ and a more confined $\theta_{\mathrm{ClCl}}$, in qualitative agreement with available measurements.\cite{Zhao2025HClDimer}

Overall, treating ionization as an explicit, time-dependent driver enables a unified interpretation of channel branching and of the resulting energy- and angle-resolved fragmentation signatures. Future work will (i) probe how laser parameters (pulse duration, wavelength, etc.) affect channel branching and macroscopic observables, (ii) test sensitivity to the initial HCl-dimer geometry by sampling different starting configurations, and (iii) extend the same laser-driven, charge-resolved analysis to other hydrogen halides dimers and larger molecular clusters of three-body and beyond.

\section{acknowledgments}

\section*{SUPPLEMENTARY MATERIAL}
Supporting Information is available for this article.

\begin{acknowledgments} 
This work was supported by the National
Science Foundation (NSF) 
under Grant No. DMR-2217759. Computational resources were provided by
ACES at 
Texas A$\&$M University through allocation PHYS240167 from the 
Advanced Cyberinfrastructure Coordination Ecosystem: Services $\&$
Support (ACCESS) program, 
supported by NSF grants 2138259, 2138286, 2138307, 2137603, and 2138296.
\end{acknowledgments}

\section*{Data Availability Statement}
The data that support the findings of this study are available
from the corresponding author upon reasonable request.

%

%

\end{document}